\documentclass[final,1p,times,twocolumn]{elsarticle}
\usepackage{epsfig}

\usepackage{amssymb}
\usepackage{amsthm}
\usepackage{amsmath}
\usepackage{amsbsy}
\journal{arXiv.org}

\begin{document}
% \linenumbers
\begin{frontmatter}



% \title{Interferometric Methods for Ultra-Wide Band Radiometry}
%\title{Interferometric Methods for Ultra-Wide Band Impulsive Signals from Ultra-high Energy Particle Showers}
\title{An Interferometric Analysis Method for Radio Impulses from Ultra-high Energy Particle Showers}


\author[label1,label10]{A. Romero-Wolf}
\author[label2,label11]{S. Hoover}
\author[label2,label12]{A.G. Vieregg}
\author[label1]{P. W. Gorham} 
\author[label4]{P. Allison}
\author[label3]{S. W. Barwick}
\author[label4]{B. M. Baughman}
\author[label4]{J. J. Beatty}
\author[label2]{K. Belov} 
\author[label5]{D. Z. Besson} 
\author[label6]{S. Bevan} 
\author[label7]{W. R. Binns} 
\author[label8]{C. Chen} 
\author[label8]{P. Chen} 
\author[label9]{J. M. Clem} 
\author[label4]{A. Connolly}
\author[label5]{M. Detrixhe} 
\author[label9]{D. De Marco} 
\author[label7]{P. F. Dowkontt} 
\author[label1]{M. DuVernois} 
\author[label3]{D. Goldstein} 
\author[label4]{E. W. Grashorn} 
\author[label1]{B. Hill} 
\author[label10]{M. Huang}
\author[label7]{M. H. Israel}
\author[label9]{A. Javaid}
\author[label1]{J. Kowalski}
\author[label1]{J. Learned}
\author[label10]{K. M. Liewer}
\author[label1]{S. Matsuno}
\author[label4]{B. C. Mercurio}
\author[label1]{C. Miki} 
\author[label6]{M. Mottram}
\author[label3,label8]{J. Nam}
\author[label10]{C. J. Naudet}
\author[label6]{R. J. Nichol}
\author[label4]{K. Palladino}
\author[label1]{L. Ruckman} 
\author[label2]{D. Saltzberg}
\author[label9]{D. Seckel}
\author[label8]{R. Y. Shang}
\author[label5]{J. Stockham}
\author[label5]{M. Stockham}
\author[label1]{G. S. Varner}
\author[label8]{and Y. Wang}

\address[label1]{Dept. of Physics and Astronomy, University of Hawai'i at Manoa, Honolulu, HI 96822.}
\address[label2]{Dept. of Physics and Astronomy, University of California, Los Angeles, CA 90095.}
\address[label3]{Department of Physics, University of California, Irvine, California 92697, USA}
\address[label4]{Department of Physics and Center for Cosmological and Astro Particle Physics, Ohio State University, Columbus, Ohio 43210, USA}
\address[label5]{Department of Physics and Astronomy, University of Kansas, Lawrence, Kansas 66045, USA}
\address[label6]{Department of Physics and Astronomy, University College London, London, United Kingdom}
\address[label7]{Department of Physics, Washington University in St. Louis, Missouri 63130, USA}
\address[label8]{Department of Physics, National Taiwan University, Taipei, Taiwan 10617}
\address[label9]{Department of Physics, University of Delaware, Newark, Delaware 19716, USA}
\address[label10]{Jet Propulsion Laboratory, California Institute of Technology, Pasadena, California 91109, USA}
\address[label11]{Kavli Institute for Cosmological Physics and Enrico Fermi Institute, University of Chicago, Chicago, Illinois 60637, USA}
\address[label12]{Harvard-Smithsonian Center for Astrophysics, Cambridge, MA 02138, USA}

% \begin{enumerate}
% \item{Check references and figures.}
% \item{Edit solar imaging additions.}
% \item{Add figures relevant to reconstruction of point sources.}
% \end{enumerate}

\begin{abstract}
We present an interferometric technique for the reconstruction of ultra-wide band impulsive signals from point sources. This highly sensitive method was developed for the search for ultra-high energy neutrinos with the ANITA experiment but is fully generalizable to any antenna array detecting radio impulsive events. Applications of the interferometric method include event reconstruction, thermal noise and anthropogenic background rejection, and solar imaging for calibrations. We illustrate this technique with applications from the analysis of the ANITA-I and ANITA-II data in the 200-1200 MHz band. 
 
\end{abstract}


\end{frontmatter}

% \linenumbers

\section{Introduction}

In the last decades there has been an increased interest in using radio frequency (RF) instrumentation for the detection of ultra-high energy (UHE) $>10^{18}$~eV neutrinos and cosmic rays. Russian-Armenian physicist Gyurgen Askaryan \cite{askaryan_1962} predicted that high energy particle showers produced in dense dielectric media would result in impulsive coherent Cherenkov radiation at radio frequencies. The emission was experimentally confirmed for the first time in 2001 using showers induced by high energy photons in silica sand~\cite{saltzberg_2001}. The results are consistent with modern particle shower and radio emission simulations \cite{ZHS92} and this effect has since been observed in salt~\cite{gorham_2005} and ice~\cite{gorham_2007}. Several experiments exploit this technique in the search for UHE neutrinos using antennas buried in ice~\cite{rice_2006,ARA}, radio telescopes pointed at the Moon~\cite{GLUE, Lunaska}, or balloon-borne antenna arrays orbiting the Antarctic continent~\cite{gorham_2009a}. 

Cosmic ray extensive air showers (EAS) produce a radio impulse due to the transverse current produced by the separation of electrons and positrons resulting from interaction with the Earth's magnetic field \cite{falcke_2003, suprun_2003}. This geo-synchrotron  emission was first observed in a ground array in the 1960's \cite{jelley_1965}. Since then, there have been many observations \cite{cr_1}-\cite{cr_12}. Recently, the Antarctic Impulsive Transient Antenna (ANITA), a balloon-borne antenna array that synoptically scans the Antarctic continent in the 200-1200 MHz range, observed geo-synchrotron emission in the ultra-high energy range for the first time~\cite{hoover_2010}.

The growth of this field demands improved analysis techniques. In particular, it is expected that the first neutrino observations will be from signals that are close to the detector threshold set by thermal noise. This requires analysis techniques that are highly sensitive and can efficiently discriminate between a weak impulse and a thermal fluctuation. 

%One way to beat the thermal noise is to coherently add the signals from $N$ antennas such that the signal to noise ratio is improved by $~\sqrt{N}$. 

Interferometric methods have been widely and successfully used in radio astronomy. Radio telescopes are able to map weak sources in the sky using the correlations between signals in an antenna array. Distant sources are imaged via the relation between phase delay and source direction \cite{TMS.1986}. Additional point-spread deconvolution methods are applied to reveal high resolution brightness maps of the radio source. The most precise astrometric measurements (200~$\mu$-arcsec resolution) are obtained from the correlations of a single pair of antennas with 8,000~km separation \cite{VLBI98}.

The signals produced by UHE particle showers are rather different from those detected by radio interferometric telescopes. UHE particle shower emissions are impulsive transient events (on microsecond to nanosecond time scales) while radio astronomical sources are better described as brightness distributions that can be imaged with long exposures. Despite the differences in the nature of these signals, the fundamental ideas of radio interferometry can be applied to the detection and analysis of UHE particle shower impulsive transient events.

An interferometric approach for impulsive signals was developed for ANITA and successfully applied to the data analysis of both flights \cite{hoover_2010,gorham_2010}. The ANITA antenna array is designed to observe impulsive radio emission in the frequency range 200-1200~MHz from UHE neutrinos interacting in the Antarctic ice. Each antenna is a dual-polarized quad-ridged horn with a gain of $\sim$10~dBi and a half-power beam-width of $~45^{\circ}$. The ANITA array is cylindrically symmetric with neighboring antennas that have a center-to-center distance of 1 meter. Typical antenna separations used in the direction reconstruction are 5 meters for ANITA-I~\cite{gorham_2009a} and 7 meters for ANITA-II~\cite{gorham_2010} producing a pointing error below $1^{\circ}$.

In the interferometric imaging procedure developed for ANITA, each pair of antennas produces a fringe across the sky at an angle corresponding to the baseline direction\footnote{A baseline is the vector defined by an antenna pair.}. The fringes are then summed together resulting in an image that peaks at the source location. Such an image is named a ``dirty map'' in radio astronomical usage, and reduction of sidelobes is possible with
further image processing. However, we have found that for this type of ``pulse-phase interferometry''~\cite{gorham_2009a}, these maps are adequate since the sources of interest to ANITA are unresolved. It is also worth mentioning that ANITA does not apply this mapping to identify and characterize source structure but rather for the identification of coherent point source impulses. We also rely on the point-like characteristics of our data and delay closure to calibrate antenna positions and cable delays~\cite{gorham_2009a}.

Although there have been other impulse beam-forming results in the past, particularly the one developed for imaging the radio flashes from ultra-high energy cosmic rays (UHECRs) for the LOPES antenna array~\cite{Falcke2005}, there are some important differences with the variant developed in this paper. ANITA is a self-triggering array and does not have muon counter data for identifying UHECR signals. The interferometric techniques developed for ANITA are applied as a stand-alone technique for identifying plane wave impulses in the data and for refining the precision of the directional reconstruction while providing improved rejection of thermal noise and anthropogenic backgrounds.

This paper presents an interferometric method applicable to broadband, impulsive radio signals for antenna arrays with large fields of view and digital waveform recording capabilities. The technique is illustrated via its application to the ANITA analysis. In Section 2 the mathematical foundations of the interferometric image production  applied to radio impulses are covered. Section 3 describes the application of interferometric images to point source impulse reconstruction along with rejection of thermal noise and anthropogenic backgrounds. Section 4 presents an application of the interferometric method to identify and characterize sources that are below detection threshold but continually present. We demonstrate the technique with observations of the Sun and images of RF activity on the Antarctic continent. In Section 5 we conclude this paper and mention some future applications of the interferometric technique developed here.

\section{Interferometric Equation}
\label{sec:interf}

In this section we formulate the interferometric approach used for radio impulses. At its core, interferometry is based on combining multiple measurements of the same signal. We discuss the relation between a recorded voltage and an electric field followed by the relation between the Adding Interferometer and a Cross-Correlation Interferometer in the context of radio impulse detection. We motivate the approach used for the analysis of ANITA data. It is worth noting, that unlike typical interferometric arrays, the ANITA antennas are not all pointing in the same direction as their boresight direction varies with payload azimuth to provide a full 360$^{\circ}$ field of view coverage~\cite{gorham_2009a}. In general, this would require that the system impulse response be deconvolved prior to beam-forming. However, we argue that for the ANITA horn antennas this is not necessary. The tools described below can just as well be applied with prior deconvolution of the antenna response but not requiring this step makes their application practical.

\subsection{Relation between the electric field and receiver voltage}
%The antenna is the interface between the electric field and voltage recordings. 
An incident electric field couples to the antenna inducing surface currents that produce voltage differences in a transmission line that can be stored in a recording apparatus. Since ANITA is primarily concerned with ultra-wide band impulses it is natural to approach the problem in the time-domain. A detailed time-domain treatment of the relation between an electric field $\mathbf{E}(t)$ and the voltage $v(t)$ recorded in an experiment can be found in~\cite{Miocinovic_2006} and references therein. The relation is captured via 
\begin{equation}
v(t, \mathbf{\hat{r}})=\sqrt{\frac{Z_{L}}{Z_{0}}}(\mathbf{h}_{sys}(\mathbf{\hat{r}})\star \mathbf{E})(t)
\end{equation}
where $Z_L$ is the load impedance, $Z_0$ is the impedance of free space, and $\mathbf{\hat{r}}$ is the direction of incidence of the radiation. The effective height vector $\mathbf{h}_{sys}(\mathbf{\hat{r}},t)$ is the time-domain representation of the antenna receiver system complex impedance, and is equivalent to the antenna receiver system response to a delta-like pulse~\cite{Miocinovic_2006}. The $\star$ operator is a vector convolution defined by $(\mathbf{f}\star\mathbf{g})(t) = \int ds \mathbf{f}(s) \mathbf{\cdot} \mathbf{g}(t-s)$. We have made $\mathbf{h}_{sys}(\mathbf{\hat{r}},t)$ explicitly a function of $\mathbf{\hat{r}}$ since, in general, the effective height depends on the angle of incidence of the electric field with respect to the antenna.  The measured voltage $v$ recorded with a digitizer is, strictly speaking, a function of time only; the added dependence of $v$ on $\mathbf{\hat{r}}$ captures the fact that the frequency contents and group delay of the effective height depends on the direction of incidence of the radiation relative to the antenna.

\begin{figure}[h]
\centering
\includegraphics[width=0.7\linewidth]{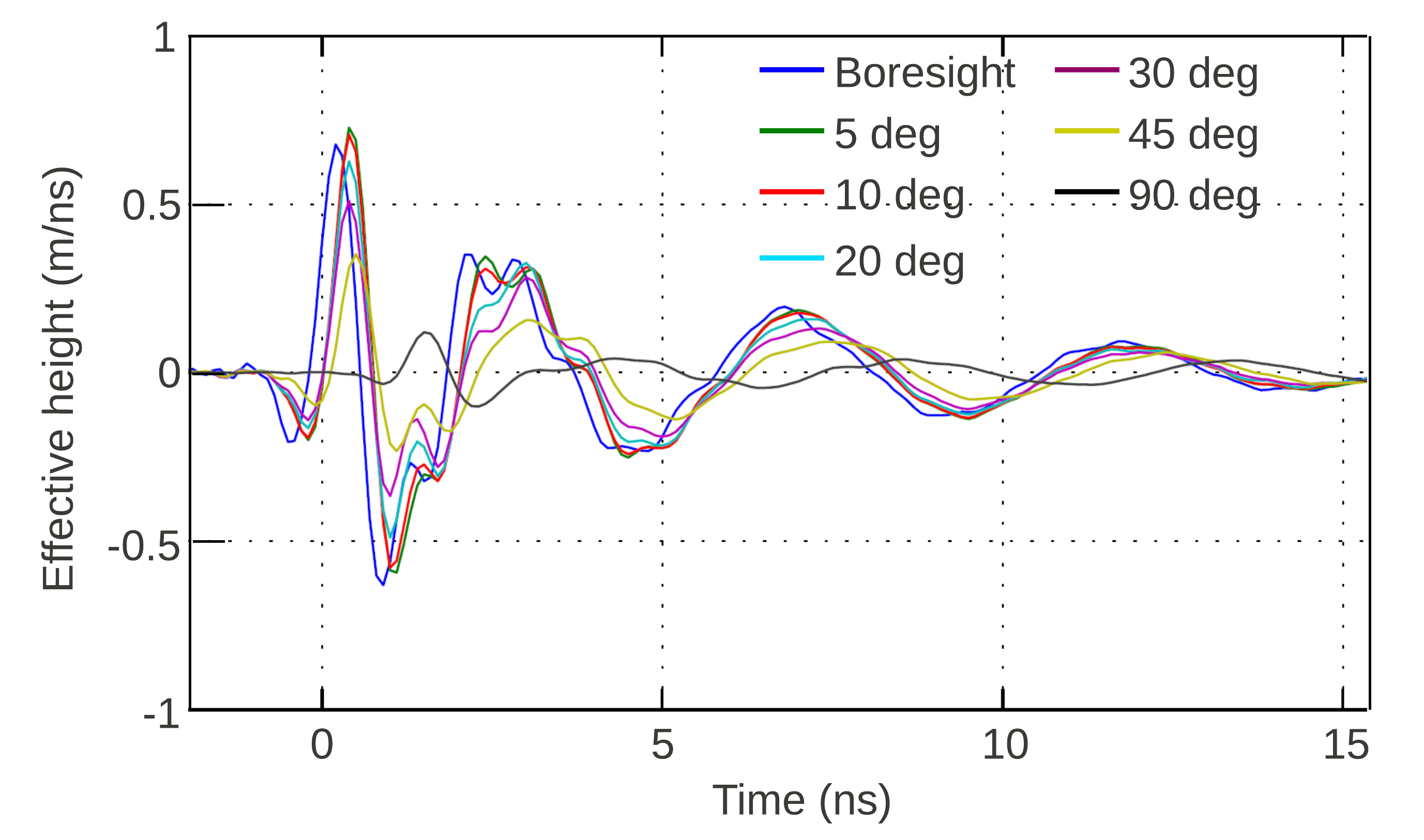}
\caption{ANITA quad-ridge horn antenna effective height for vertical polarization. The effective height is plotted for various elevation angles. The effective height near boresight has the strongest response at high frequencies showing the highest peak. As the incident electric field is moved away from boresight, the effective height loses sensitivity to the highest frequencies but retains sensitivity for the lowest frequencies. The ringing in all of these waveforms is dominated by the low frequencies ($<$ 300~MHz) and is dispersed over tens of nanoseconds. The similarity of the antenna response over a wide range of angles around boresight allows for interferometric reconstruction of source positions without the need of deconvolving the signals with the directionally dependent effective height function.} 
\label{fig:effective_height} 
\end{figure}

Figure~\ref{fig:effective_height} shows the time-domain effective height for various incidence angles for an ANITA quad-ridged horn antenna. It is important to note that Figure~\ref{fig:effective_height} shows the effective height of the antenna alone and does not include the system response of the full signal chain with cables, filters, and amplifiers. The low frequency dispersion (corresponding to the frequency range 200-300 MHz) extends for about 10~ns on the tail end of the waveform while the high frequency (300-1200 MHz) portion the signal is contained within the first few nanoseconds. Note, however, that the effective height dispersion is nearly identical for signals up to 45$^{\circ}$ away from boresight. This means that, for the purposes of correlating signals, the antenna response does not need to be corrected within this angular range. The only significant difference in the effective height function, for the various observation angles within 45$^{\circ}$ away from boresight, is the attenuation of high frequencies. This feature is equivalent to the antenna beam pattern being narrower for high frequencies and wider for lower frequencies.

\subsection{Mapping the incident signal direction using receiver voltages}
The similarity between waveforms due to the same signal in the antenna array is at the foundation of interferometry. Radio telescope interferometers typically use the pairwise cross-correlation between signals recorded at each antenna where the waveforms are delayed according to a given direction and multiplied together. Another variant is to use the Adding Interferometer where the waveforms of each antenna are delayed, summed together, and the square of the sum is integrated. Summing the waveforms, delayed according to the direction of incidence, averages down the noise while coherently adding the signal, providing an improved signal to noise ratio\footnote{We define the signal to noise ratio of an impulse by the half maximum peak to peak voltage difference divided by the root mean square of the noise.} (SNR). This summed waveform has the clear advantage of exposing a weak signal measured by a number of antennas. In this section we derive various interferometric quantities from this starting point. Although the quantities associated with the Adding and Cross-Correlation Interferometers are well known in the literature~\cite{TMS.1986}, we re-derive them here in the context of radio impulses to motivate the analysis techniques used in the next section. 
% The correlation between images is transformed into a directional map via the relationship between signal delay and the antenna array geometry. Given that a signal arrives from one particular direction, the antenna waveforms can be added coherently to enhance the signal while averaging down the noise. From this starting point we look at producing directional maps in summed waveform power, cross-correlations, and their relation. From this point we derive the coherence map used in the analysis of ANITA data.

\begin{figure}[h]
\centering
\includegraphics[width=0.5\linewidth]{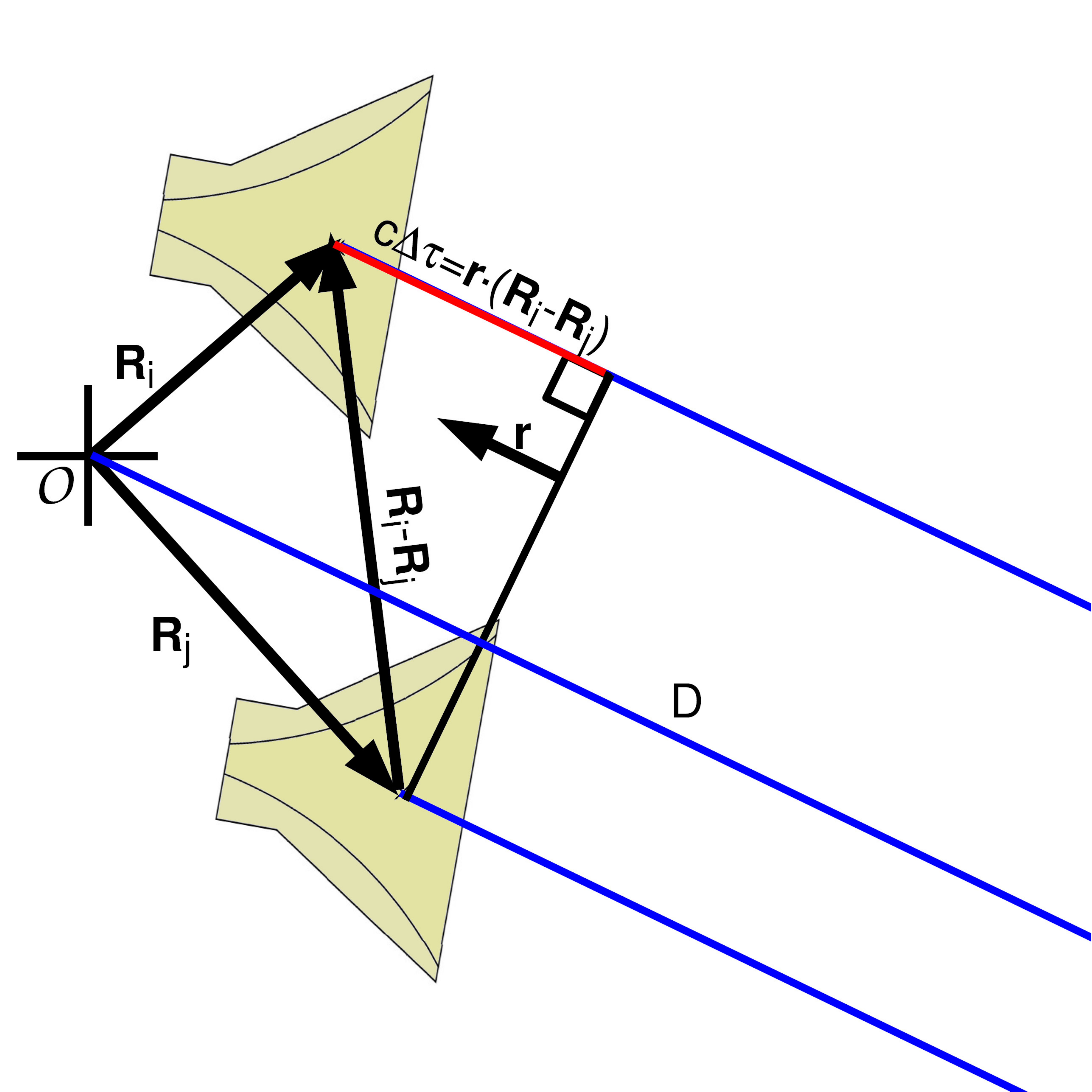}
\caption{Geometrical delay for a pair of antennas. The delayed combination of waveforms is the foundation of interferometric mapping. The schematic above shows how the delay of two signals is dependent on the geometry. Vectors $\mathbf{R}_{i}$ and $\mathbf{R}_{j}$ denote the position vectors of the two antennas. For an electric field coming from the incident direction $\mathbf{\hat{r}}$ from a distance $D$ the geometrical delay $\Delta\tau$ between the two antennas in the far-field is given by $c\Delta\tau = \mathbf{\hat{r}}\cdot(\mathbf{R}_{i} - \mathbf{R}_{j})$ where $c$ is the speed of light and  $\mathbf{R}_{i} - \mathbf{R}_{j} $ is the baseline vector. Note that the distance of the source $D$ is not present in the far-field approximation of the delay. ANITA events come from $>$100~km away for meter scale wavelengths justifying this approximation. }
\label{fig:delays} % caption for the whole figure
% \end{minipage}
\end{figure}

Let us describe the data of an array of $N_A$ antennas by a set of voltages $v_i(t)$ for each antenna $i$. For a plane wave incident from direction $\mathbf{\hat{r}}$ the delay $\tau_{i}(\mathbf{\hat{r}})$ at each antenna is given by
% Figure 1
\begin{equation}
\tau_i(\mathbf{\hat{r}})=\frac{1}{c}\left(D-\mathbf{R}_i\cdot\mathbf{\hat{r}}\right)
\label{eq:delay}
\end{equation}
where $D$ is the distance between the source and the antenna array, $\mathbf{R}_i$ is the antenna position, as shown in Figure \ref{fig:delays}, and $c$ is the speed of light. The phase aligned voltage waveforms are to be summed together to give the coherently summed waveform
\begin{equation}
V_{\Sigma}(t,\mathbf{\hat{r}})=\sum_{i=1}^{N_A}v_i(t+\tau_i(\mathbf{\hat{r}})).
\end{equation}

% Figure 2
If the voltage waveforms $v_i$ are only due to uncorrelated noise between the antennas, with the same root-mean-square (RMS) voltage $v_{rms}$, then $V_{\Sigma}$ has an RMS increased by a factor of the $\sqrt{N_A}$, regardless of the delays between them. If the voltage waveforms all contain the same plane wave signal impulse, then $V_\Sigma$ will be equal to $N_A$ times $v_i$ when the delays correspond to the direction of incidence of the signal. Thus, on average, a set of waveforms with the same noise RMS and the same signal will result in $V_\Sigma$ with an amplitude enhanced by a factor $\sqrt{N_A}$ over each individual $v_i$.

\begin{figure}[h]
\centering
\includegraphics[width=\linewidth]{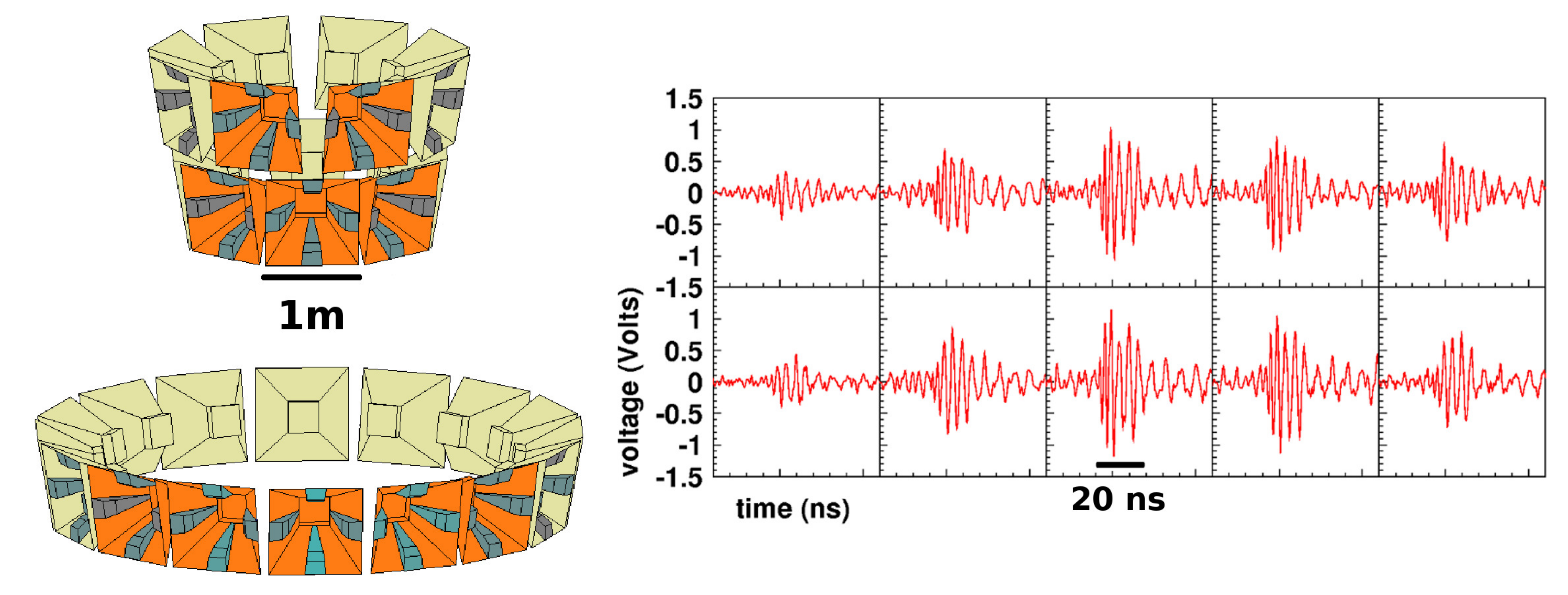}
\caption{A schematic of the ANITA horn antenna array. Each antenna is a dual polarized quad-ridged horn. The ten antennas closest to the direction of the incident impulse are highlighted with the recorded waveforms shown on the right. The signal is from a ground-based calibration pulser and only the vertically polarized channels are shown. The additional ringing in these impulses is due to a combination of the antenna response (see Figure~\ref{fig:effective_height}) in addition to the filters in the signal chain and the ringing of the transmitted impulse itself. Note that the signals are very similar within the five phi-sectors highlighted. The main difference is in the reduced high frequency response of the impulses detected away from boresight which is due to the directional dependence of the antenna effective height shown in Figure~\ref{fig:effective_height}.} 
\label{fig:payload_wfms} % caption for the whole figure
\end{figure}

\begin{figure}[h]
\centering
\includegraphics[width=0.7\linewidth]{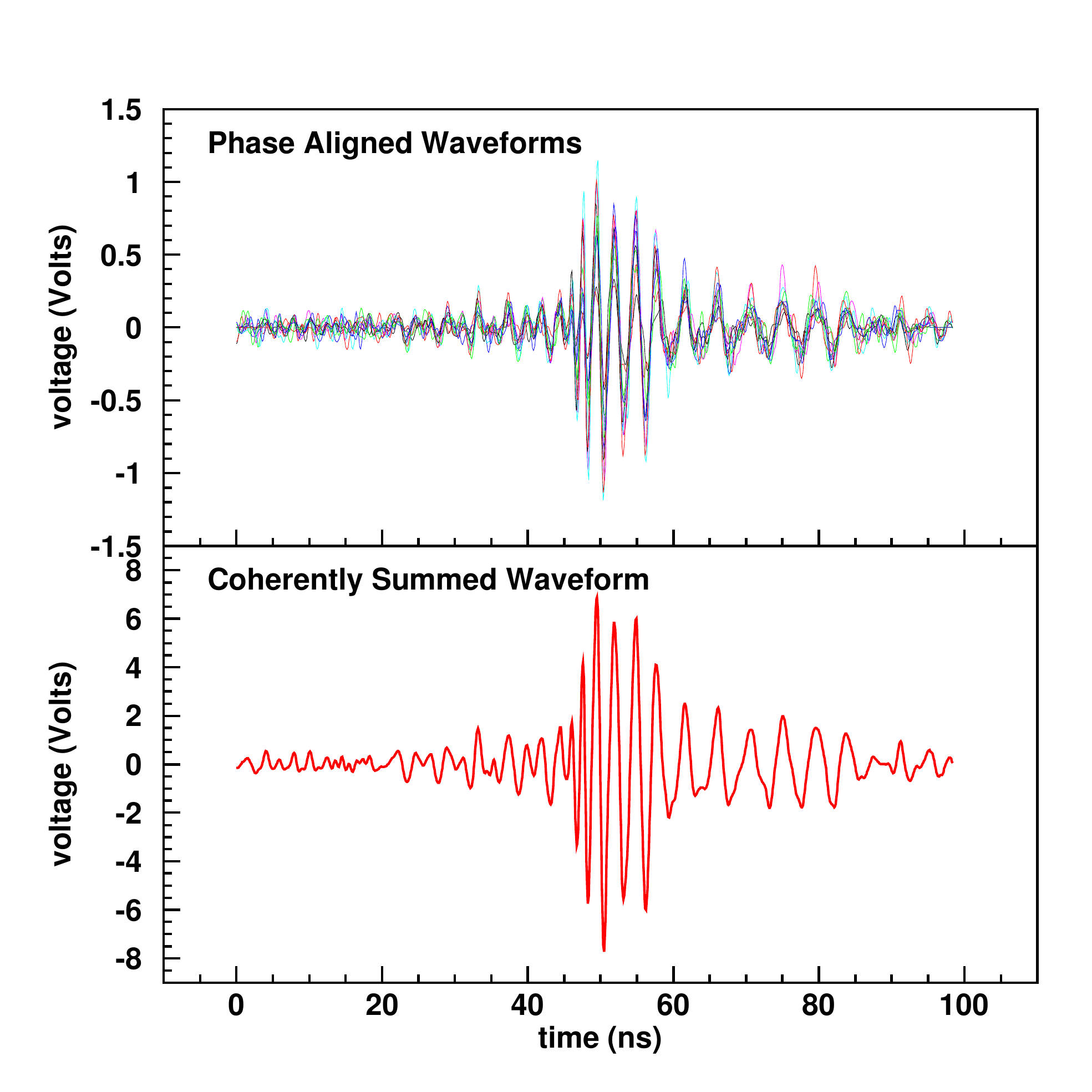}
\caption{Top: phase aligned voltage waveforms from Figure~\ref{fig:payload_wfms} according the to the known direction of incidence. Given the positions of the ground-based calibration pulser and the payload, the direction of incidence in payload coordinates can be determined with a high level of accuracy. This information is used to translate the direction of incidence to the geometrical delays expected on each antenna. Bottom: the coherent sum $V_{\Sigma}(t,\mathbf{\hat{r}})$ of all ten waveforms above. %This quantity serves the dual purpose of reducing the noise while providing a diagnostic for system calibration delay offsets.
} 
\label{fig:wfm_sum} % caption for the whole figure
\end{figure}

Figure \ref{fig:payload_wfms} shows a model of the ANITA horn antenna array with ten signals highlighted. The event shown is sent from a ground-based calibration pulser used for testing pointing reconstruction techniques. The top panel of Figure \ref{fig:wfm_sum} shows the phase (or delay) aligned waveforms using the known direction of incidence, which are summed to give the coherently summed waveform $V_{\Sigma}$ shown on the bottom panel of Figure \ref{fig:wfm_sum}.

% Figure 3
%Figure 4

One could formulate an analysis based on finding the peak of $V_\Sigma$ as a function of $\mathbf{\hat{r}}$, but in reality signals are not perfect delta functions and display some dispersion due to the antenna, the signal chain, or the neutrino induced shower itself~\cite{ARZ_2011}. For this reason, one can obtain a higher SNR measurement using the power of $V_\Sigma$ integrated over a time window $T$ relevant to the impulses of interest. 

The time averaged power of the summed receiver voltages is
\begin{equation}
% \begin{split}
 P_{\Sigma}(\mathbf{\hat{r}}) =\frac{1}{Z_L}\frac{1}{T}\int_{0}^{T}dt  \ V_{\Sigma}^2(t,\mathbf{\hat{r}}) 
% \\
% =\frac{1}{Z}\frac{1}{T}\sum_{i=1}^{N}\sum_{j=1}^{N}\int_{0}^{T}dt \  v_i(t+\tau_i(\mathbf{\hat{r}}))v_j(t+\tau_j(\mathbf{\hat{r}}))
% \end{split}
\label{eqn:sum_pow}
\end{equation}
where $T$ is the total time of the integration and $Z_L$ is the impedance of the system. The quantity $P_\Sigma(\mathbf{\hat{r}})$ is also known as the Adding Interferometer.

\begin{figure}[h]
\centering
\includegraphics[width=0.7\linewidth]{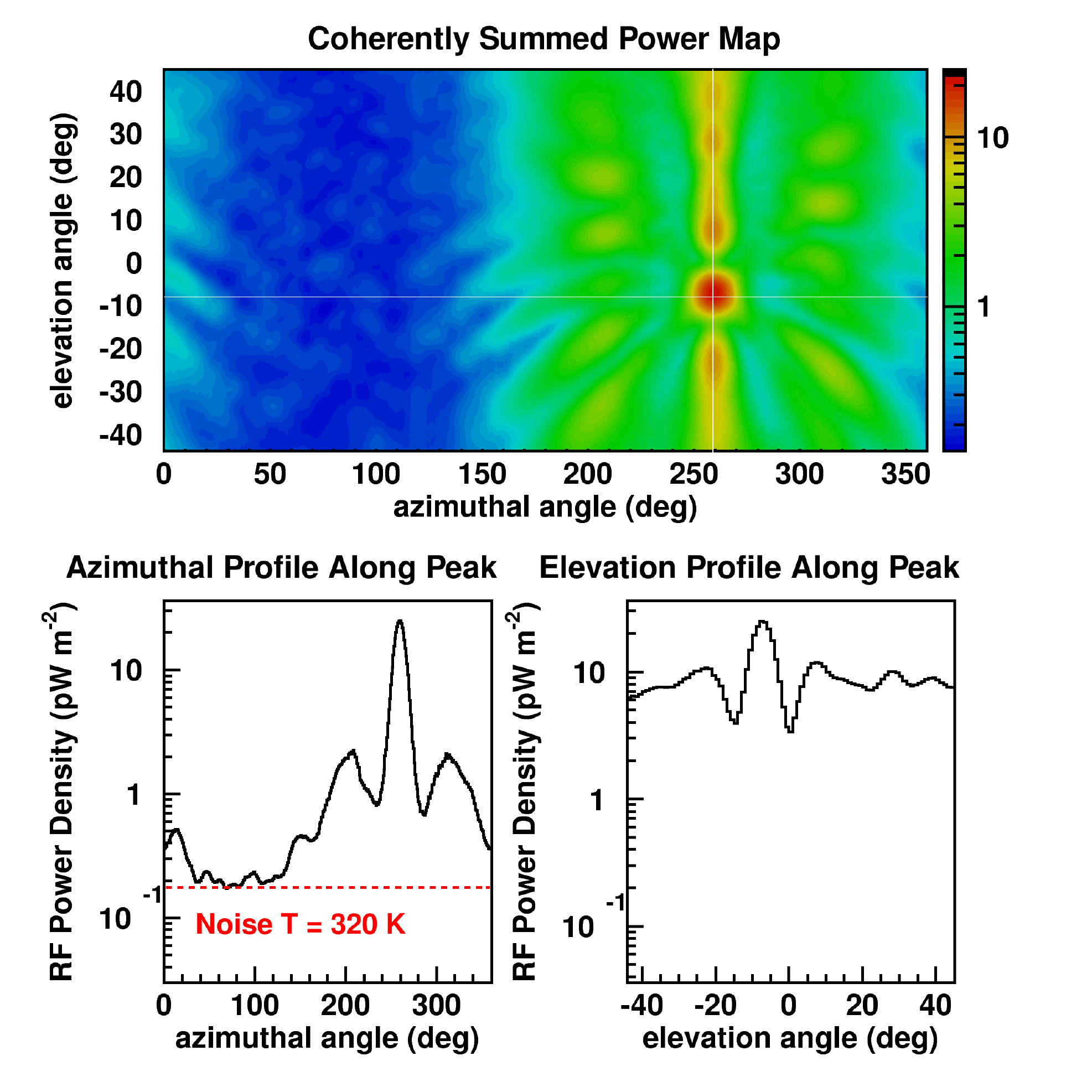}
\caption{Top: Map of the time averaged power as a function of incident direction $P_{\Sigma}(\mathbf{\hat{r}})$ for the signals shown in Figure~\ref{fig:wfm_sum}. The peak of the image corresponds to the direction of incidence of the radiation. However, rather than mapping the source structure of an object, this image identifies the location of a point source impulse. The sidelobes are due to the geometrical arrangement of the antennas.  Bottom: azimuth and elevation slices along the peak of the power map. The azimuthal slice (left) shows the summed waveform power both in the region of the peak and on the opposite side of the payload where only thermal noise is present. The power levels are consistent with system noise temperature of $320^{\circ}$~K. The elevation slice (right) has an offset due to all antennas along a phi-sector being pointed in the same direction. The minimum level of 3~pW~m$^{-2}$ is determined by the summed power of each antenna while the oscillations are due to the interference terms.} 
\label{fig:power_map} % caption for the whole figure
% \end{minipage}
\end{figure}

Figure \ref{fig:power_map} shows the power map $P_{\Sigma}(\mathbf{\hat{r}})$ for a flight calibration impulse from the ANITA-I flight. A set of ten antennas centered around each $\phi$-sector\footnote{ANITA is divided into 16 $\phi$-sectors, each consisting of a pair of antennas on top and on bottom (see left side of Figure~\ref{fig:payload_wfms}). ANITA-II has a third tier of antennas on every other $\phi$-sector}, shown in Figure \ref{fig:payload_wfms}, is used for the summed waveform as a function of direction. The power map shows the most likely direction of incidence as a large peak with sidelobes representative of the system's point spread function determined by the geometrical arrangement of the antennas and the interference pattern of the waveforms. At angles away from the direction of incidence of the impulse, the image shows the typical random pattern produced by thermal noise. Note that this approach has ignored the differences in effective height between antennas pointing in different directions. This is because the ANITA horn antenna response does not significantly vary in phase at different incident angles\footnote{Finite Difference Time Domain simulations of the impulse response of the ANITA horns found 45 ps delays between signals incident on boresight and at 22.5$^{\circ}$ away. At 2.6 Gsa/s the digitization time bin width is 384 ps.}. It also allows for faster computation which is advantageous when dealing with large data sets.
%Figure 5

In the context of mapping brightness distributions, the Adding Interferometer has the potentially undesirable feature of including the noise from each individual waveform. It also has undesirable effects when the gain and noise figures in each channel are not matched. Although $P_{\Sigma}(\mathbf{\hat{r}})$ can be a useful quantity for the analysis of radio impulses, given that the proper calibrations have been made, we can also remove some of its potentially inconvenient qualities with some mathematical manipulations. 

The time averaged power of the summed receiver voltages in Equation~\ref{eqn:sum_pow} can be expanded to give
\begin{equation}
% \begin{split}
 P_{\Sigma}(\mathbf{\hat{r}})
=\frac{1}{Z_L}\frac{1}{T}\sum_{i=1}^{N_A}\sum_{j=1}^{N_A}\int_{0}^{T}dt \  v_i(t+\tau_i(\mathbf{\hat{r}}))v_j(t+\tau_j(\mathbf{\hat{r}}))
\label{eq:sum_avg_pow}
\end{equation}
where $\tau_{i}$ and $\tau_{j}$ are the delays with respect to the origin of a given coordinate system in Equation~\ref{eq:delay}. The terms on the right hand side of the equation are proportional to the cross-correlations between antenna voltages defined as
\begin{equation}
v_{i}\otimes v_{j}(\mathbf{\hat{r}})=\int_{0}^{T} \ dt \ v_i(t)v_j(t-\Delta\tau_{ij})
\label{eq:xcorr_def}
\end{equation}
where $\Delta\tau_{ij}=\tau_i-\tau_j$. Note that the term $D$, in Equation~\ref{eq:delay}, denoting the distance from the source to the array vanishes. Substituting Equation~\ref{eq:xcorr_def} into Equation~\ref{eq:sum_avg_pow}, reduces to
\begin{equation}
P_{\Sigma}(\mathbf{\hat{r}})
=\sum_{i=1}^{N}P_i
+\frac{1}{Z_L}\frac{1}{T}\sum_{i=1}^{N_A}\sum_{j\ne i}v_i\otimes v_j(\mathbf{\hat{r}})
\label{eqn:pow2}
\end{equation}
where 
\begin{equation}
P_i = \frac{1}{Z_L}\frac{1}{T}\int_{0}^{T} \ dt \ v_i^2(t)
\end{equation}
is the average power of each individual waveform, which does not depend on $\mathbf{\hat{r}}$ given that $\tau_{ii}=0$. The cross-correlation (also known as cross-power) term contains the directional information of the electric fields incident on the antenna array.

If we only keep the terms that depend on the direction of incidence $\mathbf{\hat{r}}$ in Equation~\ref{eqn:pow2}, we obtain cross-correlation map 
\begin{equation}
X(\mathbf{\hat{r}})=\sum_{i=1}^{N_A}\sum_{j<i}v_i\otimes v_j(\mathbf{\hat{r}})
\end{equation}
where the restriction $j<i$ counts each baseline once. However, the quantity $X(\mathbf{\hat{r}})$ retains undesirable features if the gains and noise figures of each channel $i$ are not matched. 

%what would the proper convolved pattern be?
% Maximization of the interference term with respect to the direction $\mathbf{\hat{r}}$ provides the most likely direction of incidence for point source. There are several ways to apply the principles mentioned above. For example one can choose to map $P_{\Sigma}$ or use only the cross-power terms. 

Another approach is to normalize the cross-correlation by the power of the waveform according to 
\begin{equation}
C_{ij}(\mathbf{\hat{r}})=\frac{v_{i}\otimes v_{j} (\mathbf{\hat{r}})}{\sqrt{\int_{0}^{T}dt|v_{i}(t)|^2}\sqrt{\int_{0}^{T}dt|v_{j}(t)|^2}}
\label{eq:coherence}
\end{equation}
which is known as the the cross-correlation coefficient or coherence function \cite{goodman_1985}. This value is bounded between a maximum value of +1 and a minimum value of -1 and quantifies the similarity between waveforms $i$ and $j$. If the waveforms are identical the cross-correlation coefficient is +1. If they are identical with a 180$^{\circ}$ phase difference then it is equal to -1. The more dissimilar the waveforms, the closer the cross-correlation coefficient is to zero.

The power sum can be written in terms of the cross-correlation coefficients as 
\begin{equation}
P_{\Sigma}(\mathbf{\hat{r}}) = \sum_{i} P_{i} + \sum_{i\neq j} \sqrt{P_iP_j}C_{ij}(\mathbf{\hat{r}}).
\end{equation}
In this sense the cross-correlation coefficients naturally quantify the interference terms in the coherent power sum. The $C_{ij}(\mathbf{\hat{r}})$ terms are not sensitive to the overall amplitude scale of the waveforms, which can vary due to thermal fluctuations. The $C_{ij}(\mathbf{\hat{r}})$ terms contain all the directional information provided by $P_{\Sigma}(\mathbf{\hat{r}})$ of the Adding Interferometer or $X(\mathbf{\hat{r}})$ of the Cross-Correlation Interferometer. 

For the ANITA analysis it was found that the cross-correlation coefficient was the best means of reconstructing the direction of a signal~\cite{gorham_2009a, gorham_2009b, hoover_2010, gorham_2010}. The attractive features are that it normalizes out overall amplitude fluctuations along with mismatched gains and noise figures. In addition, its statistical behavior is not strongly affected by the moderate use of notch filters.

Another way to produce interferometric images is to project of the cross-correlation coefficients $C_{ij}(\mathbf{\hat{r}})$ of antenna pairs onto the incident angle space. We define the coherence map of a set of waveforms as
\begin{equation}
M(\mathbf{\hat{r}}) = \frac{1}{N_B}\sum_{i = 1}^{N_A}\sum_{j < i} C_{ij}(\mathbf{\hat{r}})
\end{equation}
where the restriction $j<i$ is put in place so as not to count the contribution of each baseline twice and $N_B$ is the number of baselines formed by a set of $N_A$ antennas given by $N_B=N_A(N_A-1)/2$.
%  excluding the pairs repeated with the opposite baseline vector direction.

\begin{figure*}[b!]
\centering
\includegraphics[width=0.71\linewidth]{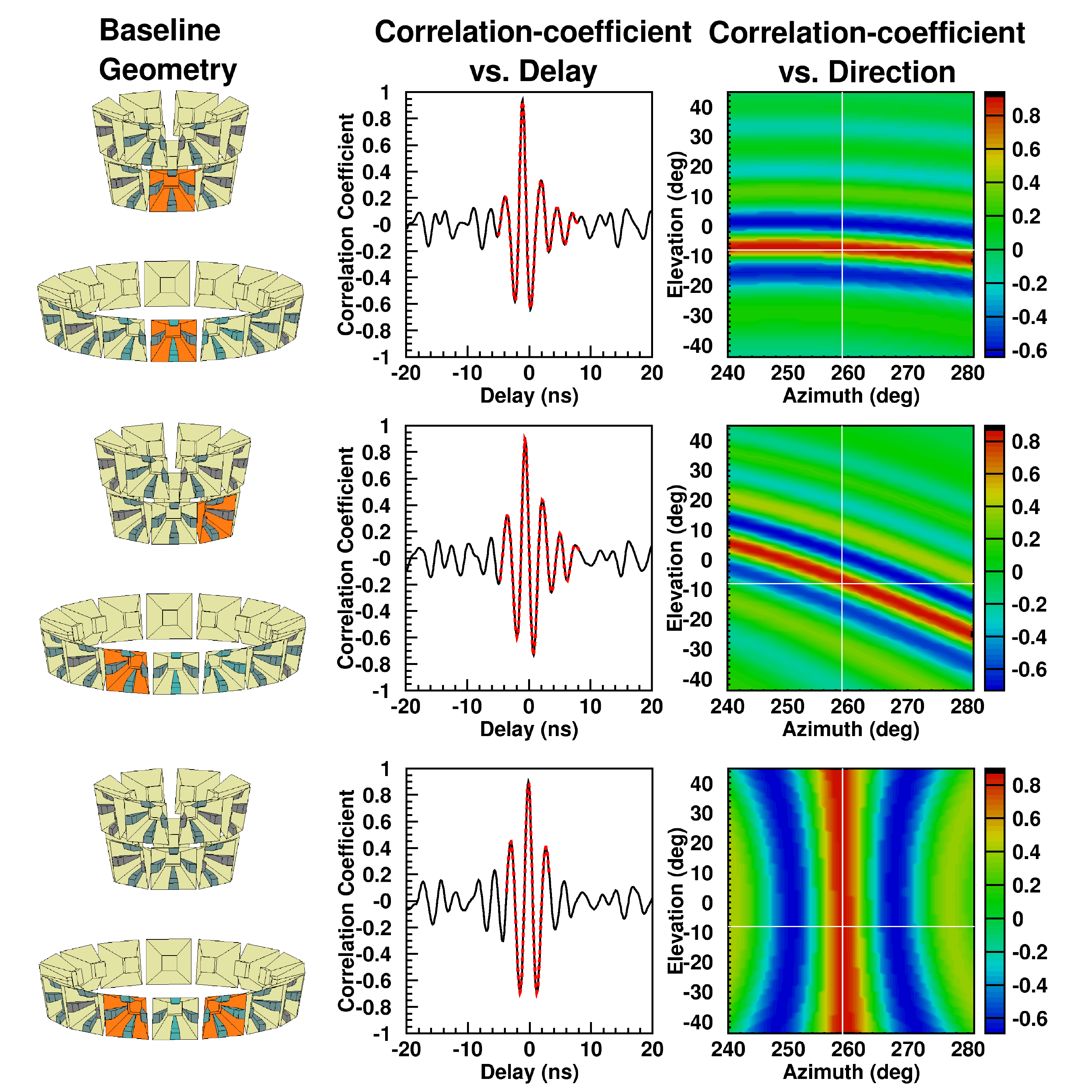}
\caption{An example of the formation of an interferometric image. From top to bottom the cross-correlation coefficient of an antenna pair is displayed. On the left, the antenna pair used for the cross correlation is highlighted. In the middle, the cross-correlation coefficient of the waveforms corresponding to the antenna pair on the left is shown as a function of time. The signals in Figure \ref{fig:payload_wfms} were used and they display a strong cross-correlation coefficient. The section of the waveform highlighted in red corresponds to delays whose geometry is constrained by the field of view of the antenna pair. On the panels to the right, the same cross-correlation coefficient is plotted as a function of incident direction in payload elevation and azimuth coordinates which is related to the delay via Equation 3. The true incident direction of the radiation lies at an elevation of -8 degrees and an azimuth of 259 degrees. Note that the directional projections of the cross-correlations all overlap at this point. } 
\label{fig:baselines} % caption for the whole figure
\end{figure*}

Figure~\ref{fig:baselines} shows the projection of a cross-correlation coefficient from signals incident on a pair of antennas for several orientations. The time-domain fringe pattern is projected to the incident direction space, in payload elevation and azimuth coordinates, and their direction is perpendicular to the baseline vector orientation. The azimuthal resolution is dominated by horizontal baselines while elevation angle resolution is dominated by the vertical baselines with contributions from the diagonal baselines. The fringe width in the incident direction plot is inversely proportional to the separation of the antennas and depends on the frequency content of the signal. 

%Figure 6

\begin{figure*}[b!]
\centering
\includegraphics[width=0.7\linewidth]{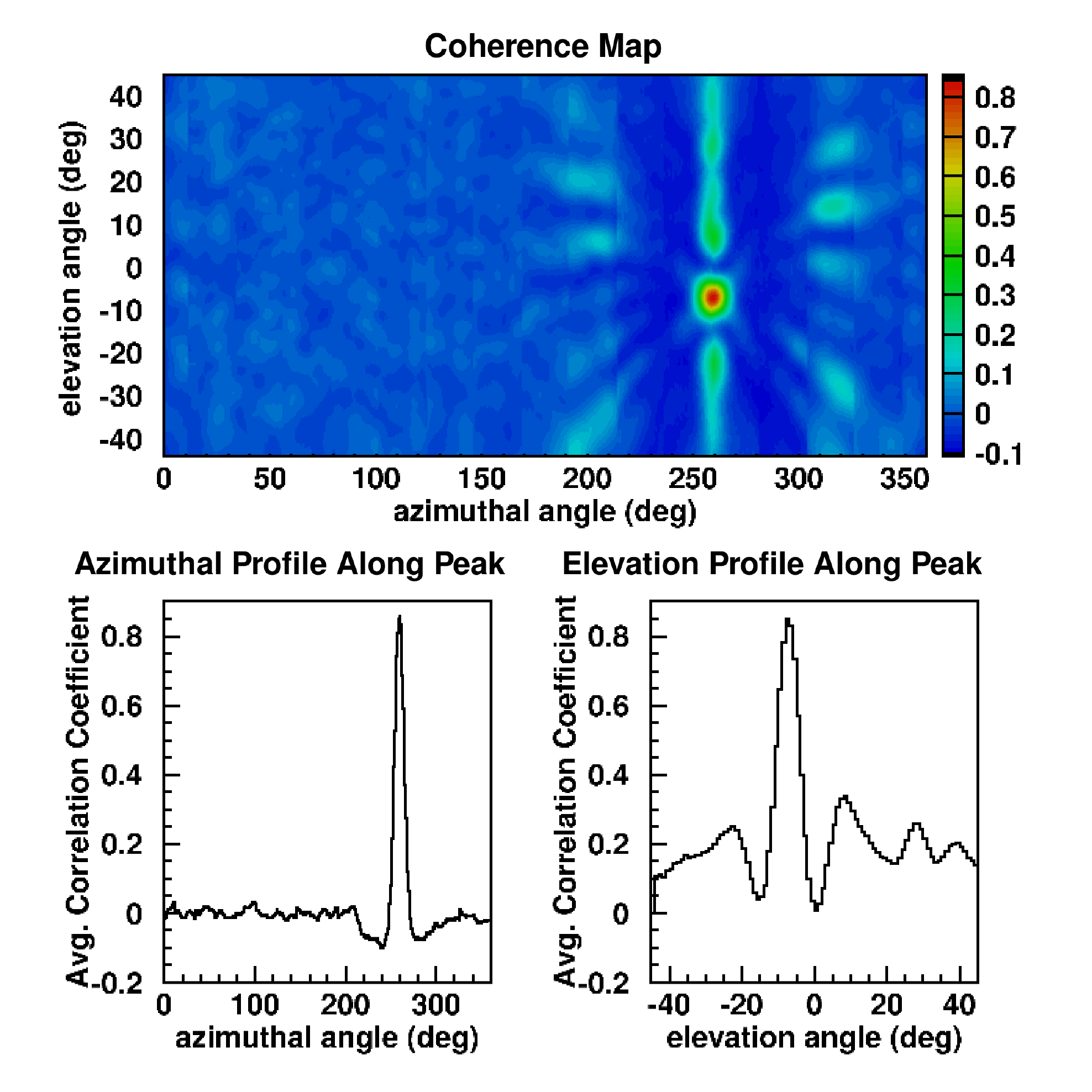}
\caption{Top: A global coherence map $M(\mathbf{\hat{r}})$ using all antenna pairs sharing the same field of view from Figure \ref{fig:baselines}. The full field of view image is a composite of coherence maps $M(\mathbf{\hat{r}})$ with an azimuthal span of $22.5^\circ$ centered around each $\phi$-sector (see text for details).  The azimuthal slice along the peak (bottom left) has very small values of $M(\mathbf{\hat{r}})$ away from the peak since all the antennas forming the image are seeing thermal noise. The elevation slice along the peak (bottom right), however, shows a larger degree of cross-correlation away from the peak. This is due to the fact that all antennas are pointed in the same direction. The formation of this image is the ``dirty map" of radio astronomy.  } 
\label{fig:interf} % caption for the whole figure
% \end{minipage}
\end{figure*}

An example of the image formed by the coherence map $M(\mathbf{\hat{r}})$ is shown in Figure~\ref{fig:interf}. The image, formed by the superposition of fringes oriented in various directions, peaks in the direction of incidence of the impulse. Note that although the individual cross-correlation coefficients $C_{ij}(\mathbf{\hat{r}})$ have sidelobes comparable to the true direction of incidence, their superposition produces a sharp peak, which greatly reduces the possibility of mis-reconstruction and increases the ability to reconstruct the direction of noisy signals. 
% Figure 7

In general, a windowing strategy is necessary to produce a coherence map $M(\mathbf{\hat{r}})$ and the choice depends on the properties of the antennas as well as the geometric configuration of the array. In the case of ANITA, the coherence map $M(\mathbf{\hat{r}})$, as shown in Figure~\ref{fig:interf} and subsequent figures, is based on the symmetry of the array as well as the antenna beam width. The global image is composed of the stitching of 16 images, each with an azimuthal field of view of $22.5^{\circ}$, centered around each $\phi$-sector of the payload. All the antennas from four adjacent $\phi$-sectors ( 5 $\phi$-sectors in total) are used in the formation of the image. %With this setup, an antenna signal that is more than $56.25^{\circ}$ away from boresight is never considered in the reconstruction of an event. 

\begin{figure*}[b!]
\centering
\includegraphics[width=\linewidth]{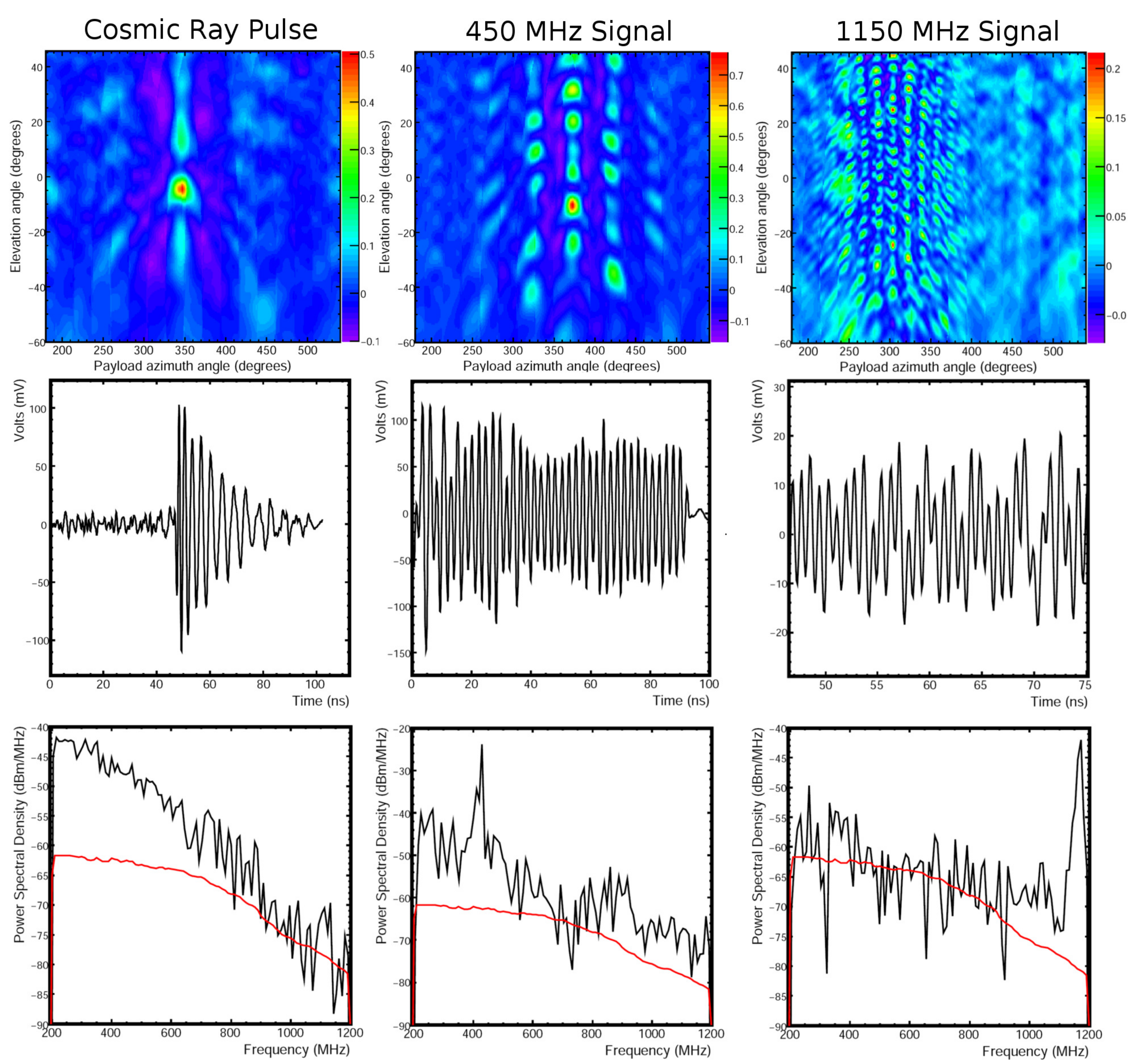}
\caption{Examples of coherence maps $M(\mathbf{\hat{r}})$ for different types of signals. From left to right a cosmic ray impulse, a 450 MHz carrier wave (CW), and a 1150 MHz CW signal are shown. In the middle row from the top, the coherently summed waveforms corresponding to the peak of the image are shown. The cosmic ray impulse shows the dispersion characteristic of the ANITA signal chain at low frequencies. The carrier waves display a strong sinusoidal behavior. On the bottom plots the power spectral densities are shown for the direction of the peak of the coherence map. The cosmic ray impulse is broadband with a falling spectrum while the carrier wave signals are strongly peaked at a single frequency. The thermal noise power spectral density is shown in red.} 
\label{fig:various_coherence_maps} % caption for the whole figure
\end{figure*}

Figure~\ref{fig:various_coherence_maps} provides examples of the coherence map $M(\mathbf{\hat{r}})$, coherent waveform sum $V_{\Sigma}(t,\mathbf{\hat{r}})$ in the direction of the peak of $M(\mathbf{\hat{r}})$, and its power spectrum for signals from the ANITA-I flight. The cosmic ray signal, shown on the left, is strong with a clearly defined peak on the coherence map. Although the cosmic ray signal is highly impulsive~\cite{hoover_2010}, the coherent waveform sum displays various oscillations due to the low frequency contents of the signal, where the ANITA impulse response is dispersive (see Section 2.1).  In the center of Figure~\ref{fig:various_coherence_maps}, a 450~MHz carrier-wave (CW) signal displays a coherence map with multiple lobes of comparable amplitude. The 1150~MHz signal, shown on the right, displays many more lobes of smaller angular size on the coherence map.
% Figure 8

In this section we have derived the Adding Interferometer equation for $P_{\Sigma}(\mathbf{\hat{r}})$ and Cross-Correlation Interferometer equations for $X(\mathbf{\hat{r}})$ and $M(\mathbf{\hat{r}})$ from the coherent waveform sum $V_{\Sigma}(t,\mathbf{\hat{r}})$. Although we primarily use $M(\mathbf{\hat{r}})$ for the ANITA analysis, this is not to say that $X(\mathbf{\hat{r}})$, $P_{\Sigma}(\mathbf{\hat{r}})$, and $V_{\Sigma}(t,\mathbf{\hat{r}})$ do not have advantages for certain applications where it makes sense to use them. As we will show in the next section, the ANITA analysis relies heavily on the combination of $M(\mathbf{\hat{r}})$ with the peak of $V_{\Sigma}(t,\mathbf{\hat{r}})$. With the relations between the interferometric quantities presented in this section being well understood, an analysis can be tailored to use each quantity, or combinations of them, as needed for the desired application.

%%%%%%%%%%%%%%%%%%%%%%%%%%%%%%%%%%%%%%%%%%%%%%%%%%%%%%%%%%
\section{Reconstruction of Impulsive Point Sources}
\label{sec:dir_reconstruc}

The coherence map $M(\mathbf{\hat{r}})$ is a powerful tool not only because it allows for the identification of a point source direction but also because it provides an efficient means to reject thermal noise and weak sources of interference. The coherence map can be combined with the coherent waveform sum $V_{\Sigma}(t,\mathbf{\hat{r}})$ and other related quantities from the previous section to provide for more sensitive analysis tools. In this section we discuss point source reconstruction and background rejection. We illustrate the application of the tools developed in this paper with examples from the ANITA analysis.

\subsection{Event Reconstruction}
The primary means of reconstructing a signal is to identify the peak of the coherence map $M(\mathbf{\hat{r}})$. The coherent waveform sum $V_{\Sigma}(t,\mathbf{\hat{r}})$ provides a means of inspection for the quality of the reconstruction. One can compare the peak of the coherence map to the peak of $V_{\Sigma}(t,\mathbf{\hat{r}})$ or $P_{\Sigma}(\mathbf{\hat{r}})$ as an additional check of the correctness and fidelity of the reconstruction. The individual waveforms $v_i(t)$ should appear aligned, as was shown in Figure~\ref{fig:payload_wfms}.

For an interferometric image, such as $M(\mathbf{\hat{r}})$, the resolution of a point source is proportional to $\lambda/R$, as prescribed by the Rayleigh criterion, where $\lambda$ is the wavelength of the radiation and $R$ is the separation between antennas. It is important to distinguish between the resolution of the image, which is the width of the peak, and the reconstruction error, which is the statistical scatter on the location of the peak after repeated measurements. In the case of wideband radio impulses the errors can be significantly better than the resolution estimate provides. The factors that affect the error of the location of the peak, besides the ratio $\lambda/R$, are the SNR, the bandwidth of the impulse, and the number of antennas observing the signal. 

For uncorrelated noise, the error on the source direction is inversely proportional to SNR (see Appendix A). The fact that a wideband impulse does not reside in a single frequency but rather over many bands requires some care in relating the resolution (proportional to $\lambda/R$) to the signal to noise ratio. The relation between the SNR of an impulse and the signal-to-noise ratio of its spectral components depends on the spectral shape of the impulse and the noise background.  For a digitizer with sampling frequency $f_s$ the frequency resolution is $\Delta f=f_s/N_s$ where $N_s$ is the number of samples recorded. For a signal with bandwidth $B$, the number of independent frequency measurements made on the signal is $N_f = B/\Delta f$. For a non-dispersive signal with a constant spectral signal to noise ratio (where the signal Fourier amplitude is a constant multiple of the average thermal noise Fourier amplitude for a given frequency range), the time-domain impulse SNR is related to the signal-to-noise ratio of the individual spectral ($snr$) amplitudes by $SNR=snr\sqrt{B/\Delta f}$, where $B$ is the bandwidth of the impulse. In this illustrative case, the error on the direction of a wideband impulse is proportional to the $\lambda/R$, where $\lambda$ is the central wavelength and inversely proportional to $snr\sqrt{B/\Delta f}$. See Appendix A for a more rigorous derivation.

The improvement in directional reconstruction errors can be estimated by counting the number of independent measurements contributing to the result. In the paragraph above we discussed the contribution of the number of bands $B/\Delta f$. For an array with $N_A$ antennas, there are $N_A-1$ independent baselines\footnote{The total number of baselines $N_A(N_A-1)/2$ can be represented as linear combinations of a subset of $N_A-1$ linearly independent baseline vectors.}. The number of independent measurements contributing to the directional reconstruction of the signal is the number of independent baselines times the number of independent frequency measurements. For the ANITA antenna geometry, the directional reconstruction error improves approximately by a factor of $\sqrt{(N_A-1)}$ over the image resolution estimate proportional to $\lambda/R$. See Appendix A for a more rigorous estimation of the point source impulsive reconstruction errors.

For the ANITA analysis, the coherence map $M(\mathbf{\hat{r}})$ is calculated with $1^{\circ}\times1^{\circ}$ pixels. The peak widths of impulsive events for ANITA are typically $\sim3^{\circ}$ for elevation and $\sim10^{\circ}$ for azimuth (see Figure~\ref{fig:interf}) depending  on the frequency contents of the impulse (see Figure~\ref{fig:various_coherence_maps}). The peak widths are consistent with the $\lambda/R$ resolution estimate for the frequencies and baseline lengths involved in making the image. When testing the angular reconstruction errors for point sources with a calibration pulser, the statistical errors are measured to be $0.26^{\circ}$ in elevation and $0.56^{\circ}$ in azimuth~\cite{gorham_2010}. For ANITA, the sampling frequency $f_s=2.6$~Gsa/s and the recording window is 256 samples giving $\Delta f \sim 10$~MHz. The calibration impulse has a bandwidth $B\sim300$~MHz. Each pixel in the coherence map image uses 10 antennas. The angular error therefore improves by $\sqrt{(N_A-1)B/\Delta f}\sim16$ over the resolution, resulting in an estimated errors of $0.18^{\circ}$ in elevation and $0.5^{\circ}$ in azimuth. The elevation and azimuth errors are consistent with the measured errors of $0.26^{\circ}$ and  $0.56^{\circ}$, respectively. 

\begin{figure*}[b!]
\centering
\includegraphics[width=0.5\linewidth]{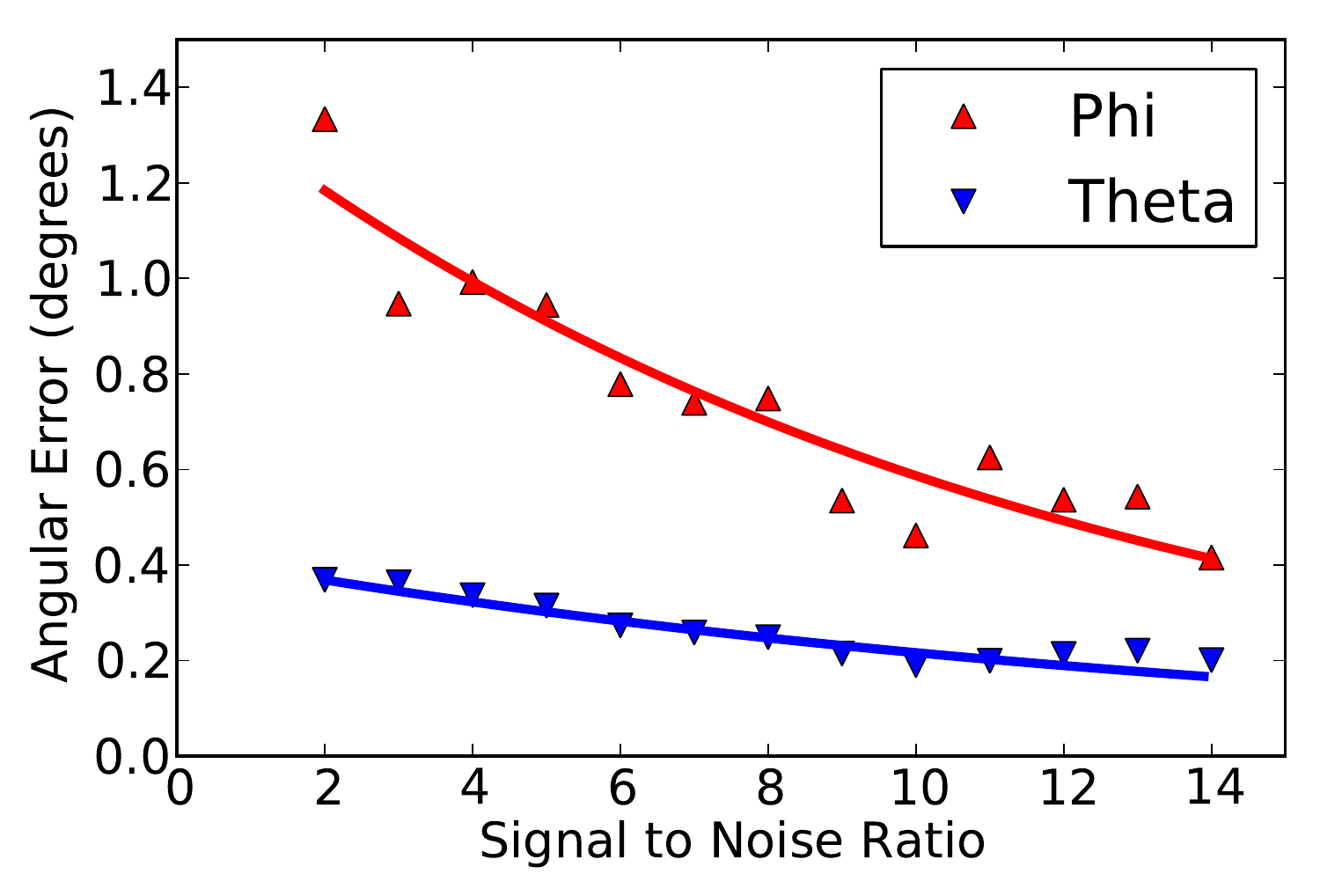}
\caption{Angular reconstruction error versus the single antenna impulse signal to noise ratio (SNR). The calibration pulser is a broadband source with most of its power at 200-400 MHz, falling off at higher frequencies. An exponential function is fitted to each trend. The angular reconstruction error in elevation (Theta) is generally better due to the fact that ANITA has long vertical baselines with antennas pointed in the same direction. The azimuthal reconstruction error (Phi) is worse due to the horizontal separation of the antennas being smaller ($\sim$1 meter) and the antennas not being all pointed in the same direction.  } 
\label{fig:resolution_vs_snr} % caption for the whole figure
\end{figure*}

Both the ANITA-I and ANITA-II data show a clear dependence on angular error with SNR. Figure~\ref{fig:resolution_vs_snr} shows the dependence of angular error on single antenna SNR for the ANITA-II pointing calibration. The quality of the angular reconstruction error can vary by a factor of 2 between the weakest and strongest signals. The comparison between the detected $SNR$, plotted in Figure~\ref{fig:resolution_vs_snr}, and the theoretical $SNR$ used in the discussion above is not straightforward due to various complications such as impulse dispersion (the signal is not a perfect delta-function impulse), the impulse signal and noise spectral $snr$ varies with frequency, and there are additional error contributions due to clock synchronization and other calibrations. However, the discussion above does provide a reasonably accurate estimate of the pointing error improvement over the image resolution.
% Figure 9

% The coherence map is not only good for reconstruction of the direction of the incoming signal; it also provides a strong handle on contamination from thermal noise and CW backgrounds when combined with the coherent waveform sum. The event quality can be further tested by looking at the coherent waveform sum $V_{\sum}$. With the event direction known from the coherence map, the individual antenna waveforms are shifted by the corresponding delays and summed together. At signal levels close to the noise, this technique has the potential to identify signals that register very weak pulses in each channel. When brought together the beam-formed signal sums coherently and the pulse surfaces out of the noise. 
% 
% The technique developed for the ANITA data analysis combines the coherence map $M(\mathbf{\hat{r}})$ and the peak value of the coherently summed waveform $V_{\sim}(\mathbf{\hat{r}})$ by first identifying the direction of the signal from the peak of $M$ and then plotting it against the peak of $V_{\sum}$ for that same direction. Figure~\ref{fig:peak_coh_vs_peak_snr} shows an example of this plot for all the vertically polarized ANITA data, along with simulated thermal noise, and simulated triggered neutrino signals. In the following subsections we discuss how this map is used to deal with thermal noise rejection and mis-reconstruction due to anthropogenic noise interference.

\subsection{Thermal Noise Rejection}
The ANITA data consist of $\gtrsim$99\% thermal noise events with potentially a few neutrino events expected near the thermal noise threshold. The ANITA analysis is searching for a small signal sample in a large thermal noise background and therefore requires a highly efficient thermal noise filter. This section will discuss how the interferometric quantities developed in this paper were used to successfully obtain a highly sensitive means of rejecting thermal noise. 

\begin{figure}[t!]
\centering
\includegraphics[width=0.75\linewidth]{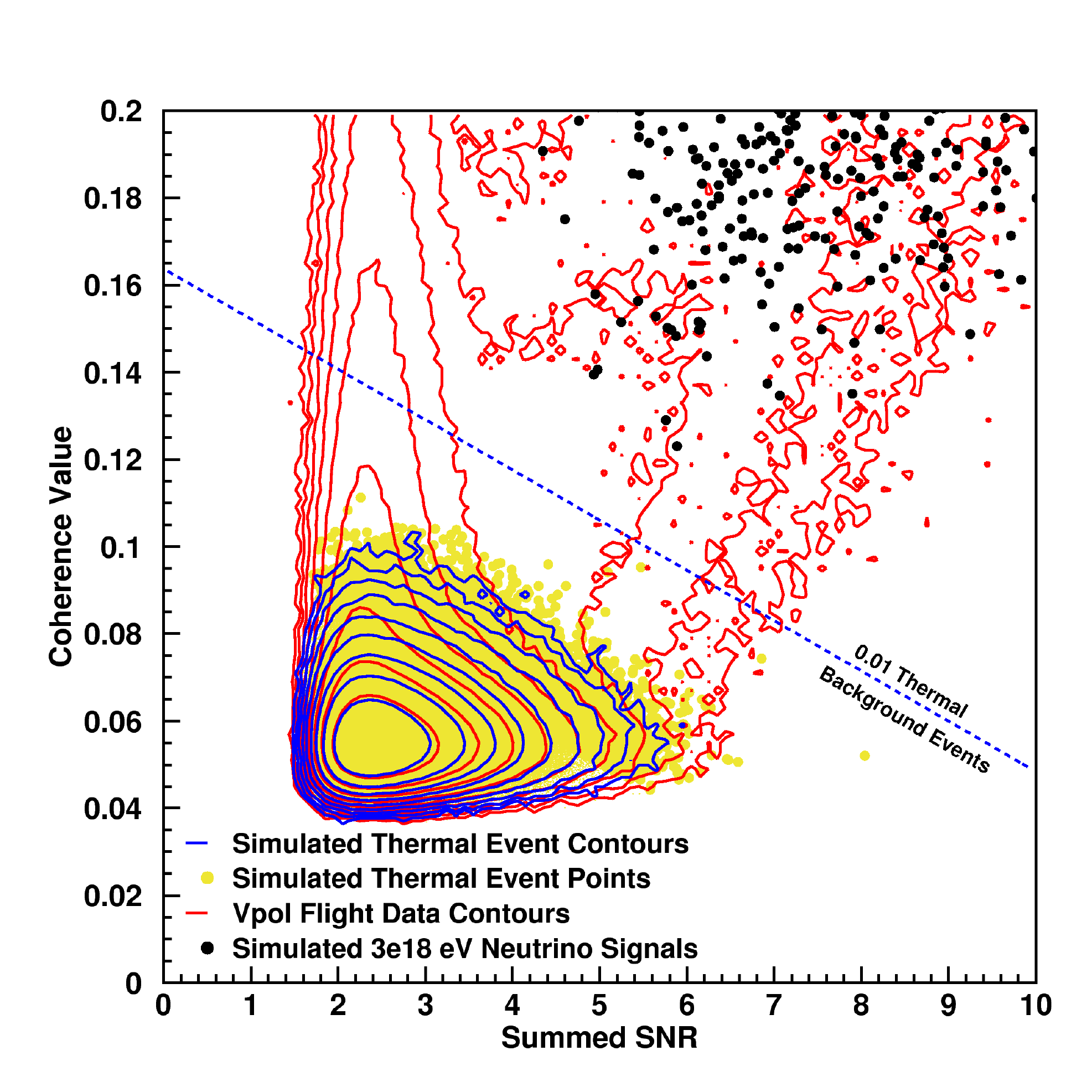}
\caption{Plot of the peak coherence map value $M(\mathbf{\hat{r}}_{max})$ versus the signal to noise ratio (SNR) of the coherently summed waveform $V_{\Sigma}(t,\mathbf{\hat{r}}_{max})$. The distribution for 8 million simulated thermal noise events (blue contours and yellow dots) shows that, at the tail of the distribution, the peak coherence $M(\mathbf{\hat{r}}_{max})$ and summed waveform SNR $V_{\Sigma}(t,\mathbf{\hat{r}}_{max})$ are anti-correlated. A contour cut is selected that admits a thermal noise event with a probability of 0.01 for the whole ANITA-I flight. The red contours are the distribution of the whole ANITA-I flight data for the vertically polarized channels. The distribution is primarily thermal and matches the simulations well. The tail to the left (high $M(\mathbf{\hat{r}}_{max})$  for low $V_{\Sigma}(t,\mathbf{\hat{r}}_{max})$ SNR) is due to residual carrier wave signals. The contours that follow a diagonal (high $M(\mathbf{\hat{r}}_{max})$ and high $V_{\Sigma}(t,\mathbf{\hat{r}}_{max})$ SNR ) are primarily due to anthropogenic signals. The black dots show triggered simulated neutrino events with energy of $3\times10^{18}$~eV. These events are distributed along the diagonal of this plot as expected.}
\label{fig:peak_coh_vs_peak_snr} % caption for the whole figure
\end{figure}

The combined use of the coherence map $M(\mathbf{\hat{r}})$ and the coherently summed waveform $V_{\Sigma}(t,\mathbf{\hat{r}})$ provide a complementary handle on thermal noise rejection. Figure~\ref{fig:peak_coh_vs_peak_snr} shows the peak coherence map value $M(\mathbf{\hat{r}}_{max})$, where $\mathbf{\hat{r}}_{max}$ is the direction of the peak, plotted against the SNR of $V_{\Sigma}(t,\mathbf{\hat{r}}_{max})$. In the case of thermal noise, these quantities become anti-correlated when either of them have a high value. If the total level of coherence fluctuates upward, the peak value of the waveform sum tends to be low. Conversely, if the noise displays a large peak, the correlation, which uses all points in the waveforms, tends to be low. This is an advantageous discriminator of impulsive signals which have a high peak coherence map value and a high summed waveform SNR. 
% Figure 10

\begin{figure}[t!]
\centering
\includegraphics[width=0.7\linewidth]{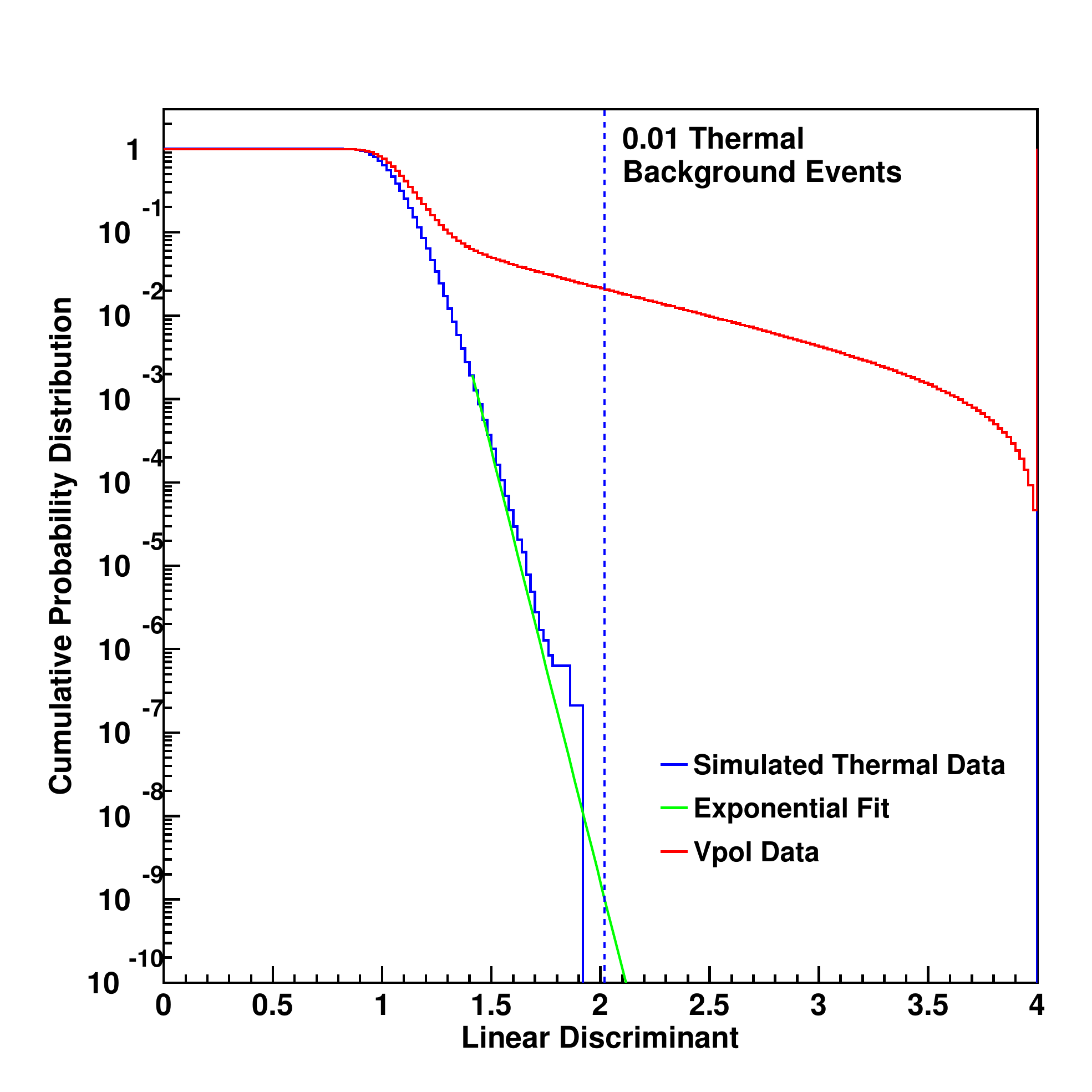}
\caption{Histogram showing the linear discriminant from Figure~\ref{fig:peak_coh_vs_peak_snr}. The blue histogram represents thermal noise simulated data for 8M events, which is the number of events recorded with ANITA-I. The green line is an exponential fit to the tail of the distribution from simulations. The cut value is set at a value corresponding to 0.01 thermal background events leaking into the analysis. The red line shows the discriminant for the ANITA-I data for vertically polarized channels.}
\label{fig:thermal_cut} % caption for the whole figure
\end{figure}

For this purpose we apply a Fisher linear discriminant, in the form of $L=y+mx$, that combines the values of the peak coherence map $y$ and coherently summed waveform SNR $x$. The slope $m$ of linear discriminant is chosen so that it is tangential to the contours shown in the thermal noise simulations in Figure~\ref{fig:peak_coh_vs_peak_snr} while $L$ shifts the overall level up and down. The distribution of the linear discriminant value $L$, shown in Figure~\ref{fig:thermal_cut}, has an exponential fall-off at the tail. The trend has been extrapolated to set a cut consistent with 0.01 thermal events passing for all of the ANITA-I data set. The value of this discriminant is shown for the all ANITA-I events in the vertically polarized channel. The black dots in Figure~\ref{fig:peak_coh_vs_peak_snr} show the distribution of 1000 simulated triggered neutrino signals~\cite{gorham_2009a} showing that the linear discriminant is a highly efficient filter of thermal noise.
% Figure 11

\subsection{Radio Frequency Interference}
Prior to the pointing analysis, the ANITA data goes through an adaptive notch filter to identify and remove CW peaks from signal spectra of the data. The most common source of mis-reconstruction is the presence of a weak CW signal that has survived the pre-filtering process. Fortunately, such signals have a tendency to produce coherence maps with multiple peaks of similar strength (see the middle panel of Figure \ref{fig:various_coherence_maps}). We are able to reject this class of events by comparing the difference of the main peak of the coherence map to the second strongest peak. This quantity provides a cut value to reduce the probability of a weak CW signal passing as an event of interest.

\begin{figure*}[t!]
\centering
\includegraphics[width=0.8\linewidth]{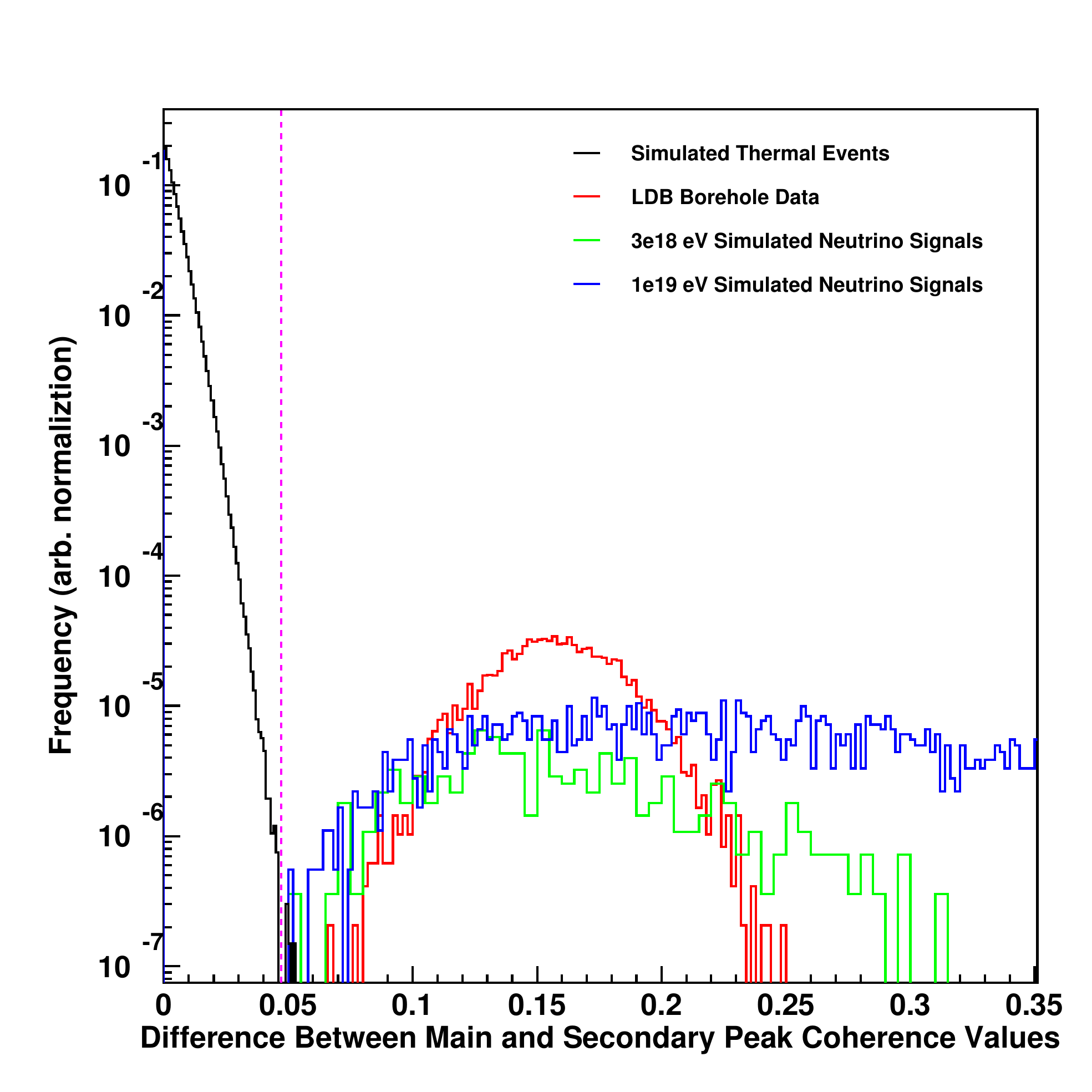}
\caption{Histogram showing the distribution for the difference between the first and second highest peaks of the coherence map $M(\mathbf{\hat{r}})$. The black histogram is for simulated thermal events and has a sharp exponential drop. The red histogram is for the calibration impulses at the Long Duration Balloon Facility in Williams Field, Antarctica. The green and blue histograms are for simulated neutrino signals with an energy of $3\times10^{18}$~eV and $10^{19}$~eV, respectively. The purpose of this cut is primarily meant to address carrier wave signals of anthropogenic origin, which tend to produce many peaks with comparable values. The magenta line allows a few thermal events to pass as signals while preserving signals of interest with 100\% efficiency. When combined with the thermal linear discriminant cut the number of thermal events passing is negligible.  }
\label{fig:peak_diff_cut} % caption for the whole figure
\end{figure*}

Anthropogenic noise is unpredictable and often involves weak CW signals in several bands. For this reason we optimize the level of the cut according to the amount of signal surviving it. Figure~\ref{fig:peak_diff_cut} shows the distribution of the difference between the first and second peak for calibration signals as well as simulated neutrino signals. The cut is set to preserve all the signals while removing a large portion of weak CW residuals. The thermal event distribution is also plotted showing a sharp exponential drop. We allow some thermal events to pass given that the linear discriminant, discussed in Section 3.2, since the linear discriminant already takes care of any potential thermal contamination in the data set.
% Figure 12

\begin{figure*}[t!]
\centering
\includegraphics[width=0.8\linewidth]{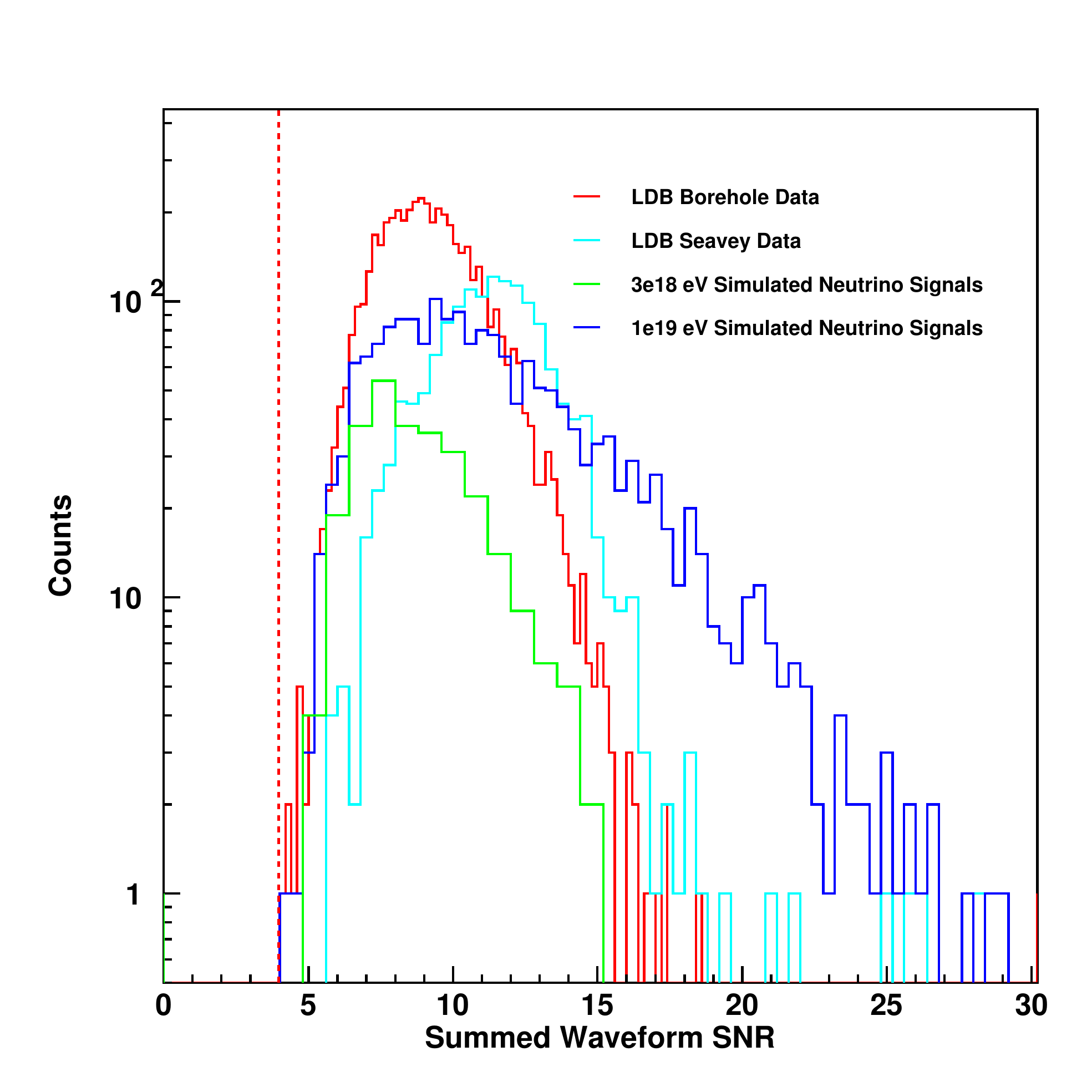}
\caption{Histogram for the coherently summed waveform $V_{\Sigma}(t,\mathbf{\hat{r}}_{max})$ signal to noise ratio distribution of signals. The red and blue histograms are signals from the borehole and ground calibration pulsers, respectively, at the Long Duration Balloon facility in Williams Field, Antarctica. The green and red histograms are the distributions for simulated neutrino events at $3\times 10^{18}$~eV and $10^{19}$~eV, respectively. The lack of events below an SNR of 4 is due to the ANITA trigger threshold. A cut on values of $SNR<4$ filters most of the CW residual signals shown in Figure~\ref{fig:peak_coh_vs_peak_snr}}
\label{fig:snr_cut} % caption for the whole figure
\end{figure*}

The effect of weak CW interference also manifests itself in the comparison of the peak coherence value $M(\mathbf{\hat{r}}_{max})$ versus the SNR of $V_{\Sigma}(t,\mathbf{\hat{r}}_{max})$ (the maximum peak-to-peak distance of the coherently summed waveform calculated in the direction of peak coherence). The population of events in the data that produce the contours on the region between an SNR of 1 and 3 and coherence values greater than 0.1, in Figure \ref{fig:peak_coh_vs_peak_snr}, is due weak CW contamination, which results in an increased coherence while not displaying a strong SNR in $V_{\Sigma}(t,\mathbf{\hat{r}}_{max})$. The distribution of coherent waveform sum SNR is shown in Figure~\ref{fig:snr_cut} for calibration signals and simulated data. Cutting data with SNR$<$4 filters out most weak CW interference while retaining signals with 100\% efficiency.

\section{Identification and Characterization of Weak Signals}
\label{sec:solar}
If a stationary  source is too weak to be detected in a single interferometric
image, the images can be stacked to provide a strong detection. This technique can be useful for characterizing known weak sources of
noise as well as identifying new ones. We demonstrate the use of averaging interferometric images with short integration times to observe the Sun. We apply a similar technique to producing RF source maps of the Antarctic continent.

\subsection{Solar Imaging}

The Sun, while a significant thermal source in ANITA's frequency range, is not distinguishable in a single interferometric image created from ANITA data. Each ANITA interferometric image contains only $\sim$100~ns of data. To detect such a source, one must combine many events by averaging the individual interferometric images together. Selection of a coordinate system in which the source of interest is stationary allows the signal to remain constant while the noise averages down, and is essential to this process. 

So far, we have plotted the coherence map $M(\mathbf{\hat{r}})$ as a function of payload elevation $\theta_p$ and azimuth  $\phi_p$. This is the coordinate system used in Figures \ref{fig:power_map}, \ref{fig:baselines}, \ref{fig:interf}, and \ref{fig:various_coherence_maps}. However, we can just as well select another reference coordinate system for the image. For the purpose of imaging the Sun, we use solar azimuth $\phi_S$, where $\phi_S=0$ corresponds to the location of the Sun. Although at South Polar latitudes the Sun changes elevation angle throughout the day, it is reasonably stationary in the 30 minute time scales used for averaging. Making the coherence map in this coordinate system reduces to recalculating the baseline delays $\tau(\theta_p, \phi_S)$ and averaging the cross-correlation coefficients for each baseline $C_{ij}(\tau(\theta_p, \phi_S))$ to produce the coherence map $M(\theta_p, \phi_S)$. Each map, using a $\sim$100~ns snapshot in the case of ANITA, will not reveal an image of the Sun. However, the average map over $N$ events
\begin{equation}
S(\delta, \phi_S) = \frac{1}{N}\sum_{ev=0}^{N}M_{ev}(\theta_p, \phi_S),
\end{equation}
where the index $ev$ runs over all events used in the average, will produce an image given that $N$ is large enough.

\begin{figure*}[t]
\centering
\includegraphics[width=0.8\linewidth]{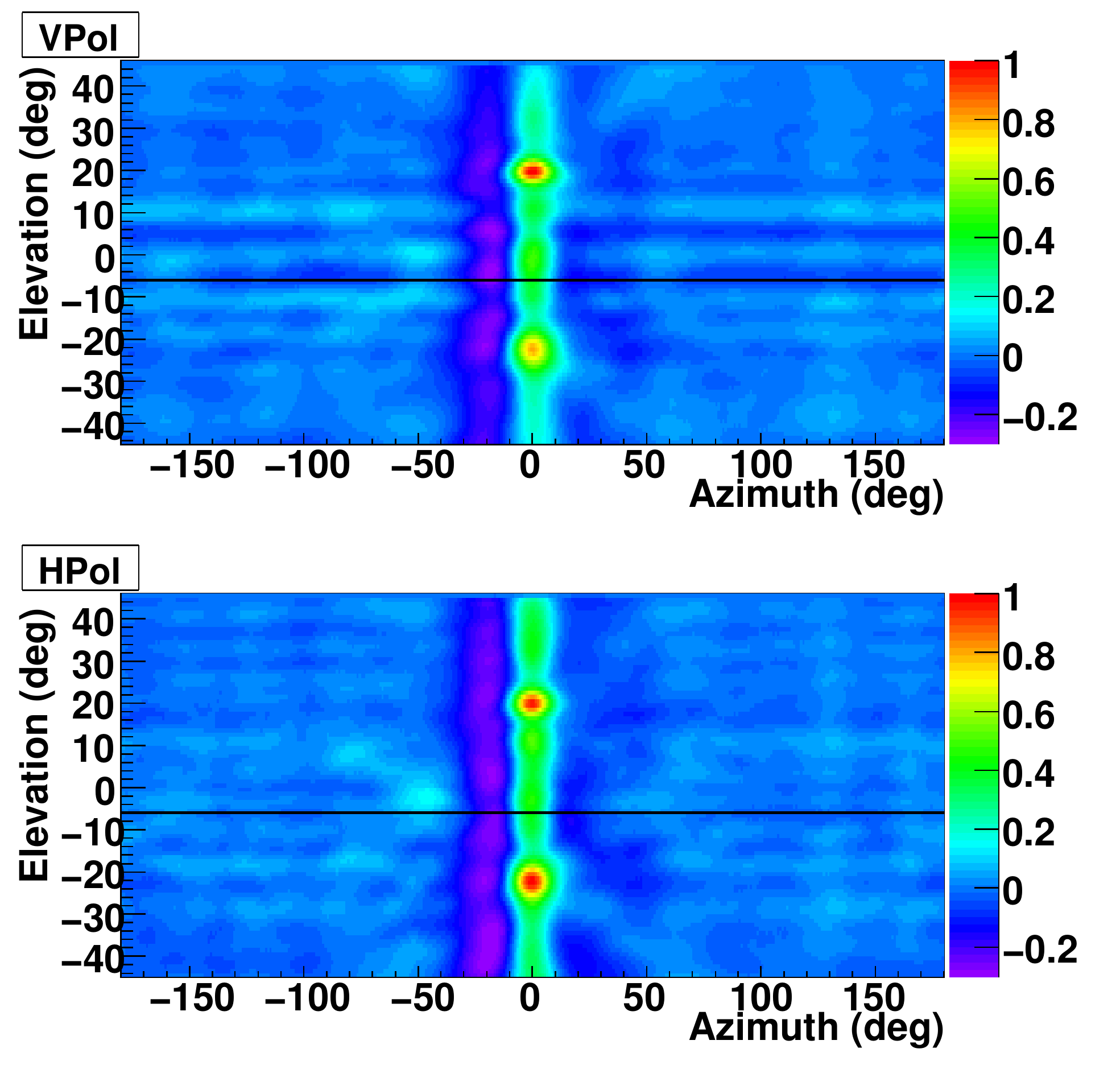}
\caption{An image of the Sun formed by averaging  $10,000$ interferometric images of ANITA events. The location of the Sun is tracked throughout the three hours used to form the image and is set as a reference point. The Sun is at $20^{\circ}$ in payload elevation angle.  The azimuth angle coordinate is centered on the expected payload azimuthal location of the Sun. The fainter peak at -20$^{\circ}$ elevation is consistent with the reflection of the Sun on the ice. The reflection point is stronger in the horizontally polarized channels as expected. The color scales in the coherence map are normalized to the peak value of the map. Features immediately surrounding the image of the Sun are the sidelobes of ANITA's point spread function. The straight bright curves, which extend across all azimuths, could be associated with the horizon (marked with a black line) as seen from ANITA's latitude of $\sim 80^{\circ}$~south at an altitude of 37~km.  }
% \caption{Coherence map of the sun. This map is obtained by averaging the interferometric image of thousands of events.} 
\label{fig:sun} % caption for the whole figure
% \end{minipage}
\end{figure*}

Figure \ref{fig:sun} shows an image of the Sun $S(\theta_p, \phi_S)$ created by averaging
together 10,000 ANITA events ($\sim$1 ms of data, recorded over a $\sim$30 minutes
period). The image created from the average amplitudes of each pixel
clearly reveals the Sun and its surrounding sidelobes. This image shows a peak corresponding to the location of the Sun at $\theta_p\sim$20$^{\circ}$. In addition, the image shows a bright spot consistent with the Sun's reflection on the ice at $\theta_p\sim$-20$^{\circ}$. The reflection is stronger in the horizontally polarized image than the vertically polarized image, as expected from the Fresnel reflection coefficients on the surface of the ice, with index of refraction $n=1.35$. Another interesting feature are the horizontal bright lines spanning the full azimuthal field of view. These lines could be due to the horizon where the noise temperature transitions from 270$^\circ$~K on the surface to 10$^\circ$~K on the sky. No conclusive evidence has been found to distinguish it from a detector artifact and the effect is currently under investigation.
% Figure 14

This technique is being further explored for a variety of applications and will be treated in detail in an upcoming publication~\cite{Besson_2013}. The Sun provides a constant source that allows us to monitor the antenna gain calibration throughout the flight. We can also monitor the surface radio reflectivity by comparing the brightness of the Sun and its reflection both in the vertical and horizontal polarizations. This comparison provides a direct measure of the index of refraction of the surface of the ice. Surface roughness estimates are important to the energy determination of reflected ultra-high energy cosmic ray air shower events~\cite{hoover_2010}. 

% \section{RF Imaging of the Antarctic Horizon}
\subsection{Man-Made RF Activity on the Antarctic Continent}
\label{sec:anthro}
Except for a small fraction of signals expected from ultra-high energy particles, ANITA records two main types of events: first, thermal noise fluctuations that trigger the system comprising $\gtrsim$99\% of the data and second, anthropogenic signals originating from Antarctic bases, field camps, traverses, and potentially aircraft comprising the remaining $\lesssim$1\% of the data. There is, however, some overlap in these events. Some anthropogenic sources can be well below ANITA's trigger threshold but still present in the thermal noise triggered data. Much like the treatment of the Sun in section 4.1, we are able to make subthreshold maps of the Antarctic continent.

For ANITA-I, a total of 8~M events were collected, making the projection of all coherence maps onto a coordinate system covering Antarctica impractical. However, we can take the peak value and direction of the coherence map $M(\mathbf{\hat{r}}_{max})$ for each event and project it onto the Antarctic continent. In the following, we describe the mapping procedure. First, we find the peak direction of the coherence map in payload coordinates $\hat{\theta}_p$, $\hat{\phi}_p$. We then transform the direction to a local East, North, Vertical (ENV) coordinate system to obtain the direction of the peak in $\hat{\theta}_{ENV}$, $\hat{\phi}_{ENV}$. This involves correcting for payload heading and attitude offsets from vertical~\cite{gorham_2009a}. The location of the payload is determined by an on-board global positioning system (GPS) unit reported in Earth-Centered Earth-Fixed (ECEF) Cartesian coordinates. The direction of the peak coherence also needs to be transformed to $\hat{\theta}_{ECEF}$, $\hat{\phi}_{ECEF}$. Given the position and direction of the peak coherence, we can propagate the ray to a location on the Antarctic continent. To do this, we bin the continent in Easting, Northing coordinates\footnote{Easting and Northing coordinates place the South Pole as the origin with the y-axis (Northing) pointed along $0^{\circ}$ longitude line and the x-axis (Easting) pointed along the $90^{\circ}$ longitudinal line.} in 10 km by 10 km squares. Using an elevation model of the continent~\cite{RadarSat} we determine the ECEF coordinates ($x_k,y_k,z_k$) for the center of each bin $k$. We then linearly propagate the ray from ($x_p,y_p,z_p$) in the direction of $\hat{\theta}_{ECEF}$, $\hat{\phi}_{ECEF}$ to find the closest bin ($x_k,y_k,z_k$) consistent with that propagation.

\begin{figure}[b!]
\centering
\includegraphics[width=0.7\linewidth]{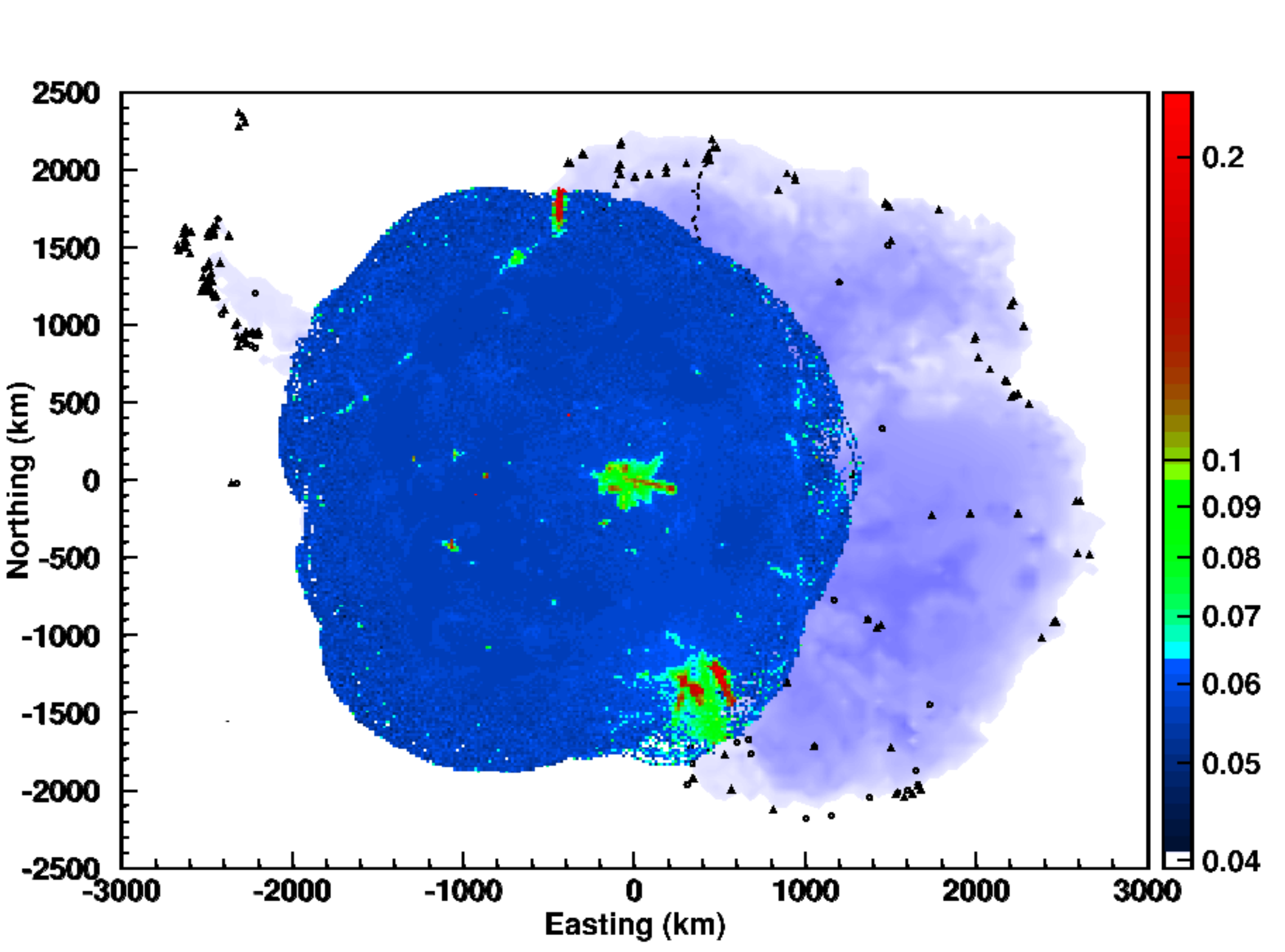}
\caption{Coherence map of the Antarctic continent for the ANITA-I flight. The image is formed by obtaining the peak value of the coherence map of each ANITA event and then projecting its direction onto the Antarctic continent.  The color scale indicates the average value of the peak of the coherence map $M(\mathbf{\hat{r}}_{max})$ for all triggered events which project onto a given bin on the map. This map has been useful for identifying the structure and location of anthropogenic backgrounds. Since no cuts are placed on the data prior to its projection onto the continent, it allows for the identification of regions of anthropogenic radiation that are below the threshold of the neutrino search cuts. Note that a large portion of the Antarctic continent (in dark blue) is radio quiet. McMurdo, on the bottom, and South Pole station, in the middle, are very strong emitters and dominate the radio anthropogenic background.} 
\label{fig:icemap_anita1} % caption for the whole figure
\end{figure}

\begin{figure}[t!]
\centering
\includegraphics[width=0.7\linewidth]{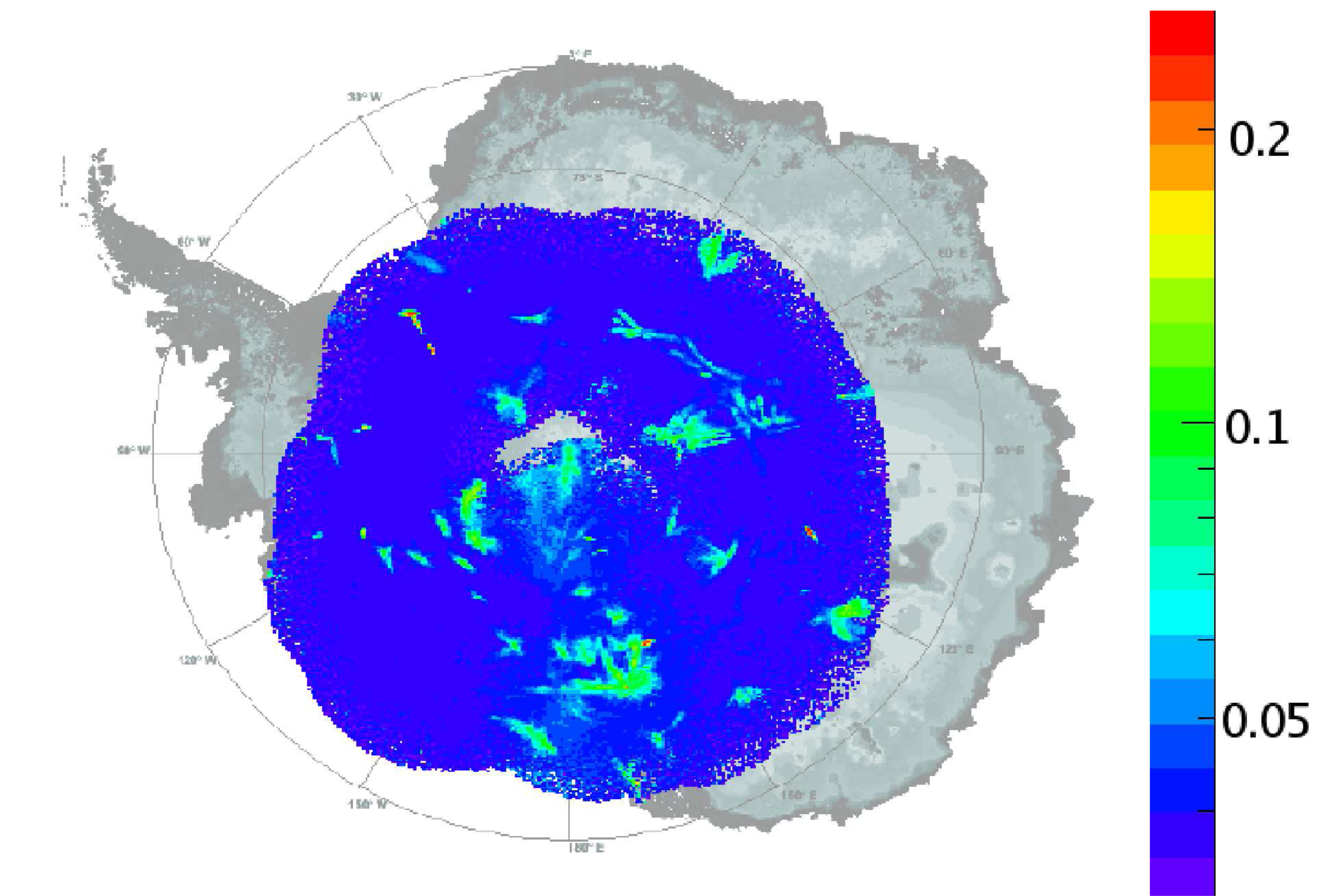}
\caption{Coherence map of the Antarctic continent for the ANITA-II flight. The image is made in the same fashion as that shown in Figure \ref{fig:icemap_anita1}. McMurdo displays the largest and strongest emission. ANITA-II did not fly over the South Pole like ANITA-I but it was still able to observe strong anthropogenic backgrounds in its vicinity. The remaining bright spots are emissions from field camps and other radio experiments.} 
\label{fig:icemap_anita2} % caption for the whole figure
\end{figure}

Figure~\ref{fig:icemap_anita1} shows the average of the peak coherence values for events projected into each bin on the Antarctic continent for vertically polarized data of all ~8~M events from ANITA-I. The blue background is consistent with the thermal noise expectation, while the colored regions indicate hot-spots of anthropogenic activity. Figure~\ref{fig:icemap_anita2} shows a similar map made using all 21.2~M events from ANITA-II~\cite{gorham_2010}. More anthropogenic hot-spots were observed with ANITA-II because of the increased exposure and sensitivity of that flight.
% Figure 15 %%%%%%%%%%%%%%%%%%%%%%%%%%%%%%%%
% Figure 16 %%%%%%%%%%%%%%%%%%%%%%%%%%%%%%%
%%%%%%%%%%%%%%%%%%%%%%%%%%%%%%%%%%%%%%%%%%%%%%%%%%%%%%%%%%

When these maps, made for all the data, are compared to the event clusters passing the thermal noise and CW filters, we find clusters of activity that do not appear in the latter population of events. This indicates that there is potential for a single anthropogenic transient event that could be detected from among the background of sub-threshold signals. We therefore exclude any events from such sites in the final analysis~\cite{gorham_2010}. It is also interesting to note that this technique provides a diagnostic for the radio quiet properties of a given site. If we look at how the events in each bin are distributed in peak coherence, we can determine whether the site is consistent with pure thermal noise or if there is some sub-threshold tail indicating weak anthropogenic activity. We have used this technique for visual inspection and we are currently looking into applying it more formally to the data analysis.

Reliable directional reconstruction of anthropogenic activity was critical to the results of the ANITA-I and ANITA-II ultra-high energy neutrino searches~\cite{gorham_2009b, gorham_2010}.  The criteria for tagging an above threshold event (an event passing the thermal noise and CW filters) as anthropogenic is any one of the following: the event clusters with other above threshold events, the event clusters with a known location of human activity regardless of whether it also clusters with other above threshold events, the event clusters with a local maximum (hot-spot) from Figure~\ref{fig:icemap_anita1} for ANITA-I or Figure~\ref{fig:icemap_anita2} for ANITA-II which may be formed by purely subthreshold signals. If any of these cases are satisfied the event is rejected.  For ANITA-I, one isolated event was identified as a man-made background event by association with a hot-spot formed solely out of subthreshold events.  Even though there is 
a significant amount of human activity on the Antarctic continent, especially as seen by ANITA-II, 
only 36\% of the continent was excluded in the ANITA-II neutrino search analysis due 
to proximity to anthropogenic events~\cite{gorham_2010}.  The blue regions shown in Figures~\ref{fig:icemap_anita1}~and~\ref{fig:icemap_anita2}, indicate quiet regions where neutrino candidate signals could be found.

\section{Outlook and Conclusions}

The interferometric techniques developed for impulsive signals are applicable not only to event reconstruction but also to filtering of thermal noise and radio interference as well as the identification of weak background sources. The methods presented are suited for antenna arrays with digital sampling capabilities, the kind of detector fit for searches of UHE particle impulsive transient radio emission. The applications discussed in this paper are only a subset of the capabilities provided by the interferometric technique and several others are currently under development.

Future applications of the Solar imaging technique include estimation of the surface roughness of the Antarctic ice by comparing the direct and reflected images. This technique can also be applied to a determination of the index of refraction and reflectivity of the Antarctic surface. The analysis of Solar reflections will also provide a measure of surface roughness on the Antarctic continent relevant to the energy determination of ultra-high energy cosmic ray events~\cite{Besson_2013}.

The interferometric principles presented in this paper are also being applied towards the development of a new trigger for the third flight of ANITA. The interferometric trigger will continuously digitize data with 3-bit resolution in real-time. The beam-forming can then be performed in real-time using an application-specific integrated circuit or a field-programmable gate array, if the power consumption allows. The development of the interferometric technique for impulsive transients applied to hardware algorithms will be treated in a future publication.

The application of interferometric imaging to impulsive signals has proved to be a powerful technique. It is by no means limited to what has been presented in this paper and it will continue to be exploited in future efforts for the radio detection of ultra-high energy particles.

\section{Acknowledgments}
We thank the National Aeronautics and Space Administration,
the National Science Foundation Office of Polar Programs,
the Department of Energy Office of Science HEP Division,
the UK Science and Technology Facilities Council, the
National Science Council in Taiwan ROC, and especially the
staff of the Columbia Scientific Balloon Facility.
A. Romero-Wolf would like to thank NASA (NESSF Grant NNX07AO05H) 
for support for this work. A.G. Vieregg would like to thank NASA (NESSF Grant 09-Astro-09F-0008) and the National Science Foundation (Grant No. ANT-110355). Copyright 2013. All rights reserved.

\pagebreak

\appendix
\section{Statistical Pointing Error Estimation}
\subsection{Probability Density Function and Likelihood Estimator}
In this appendix we estimate the pointing errors from a likelihood estimator approach. Following~\cite{goodman_1985}, the probability density function $p(a,\phi)$, for the amplitude $a$ and phase $\phi$ of a phasor $\tilde{a}=ae^{i\phi}$, resulting from the sum of a number of random contributions large enough to satisfy the central limit theorem, is
\begin{equation}
p(a,\phi)dad\phi = \frac{a}{2\pi\sigma^2}\exp\left[-\frac{a^2}{2\sigma^2}\right] dad\phi,
\end{equation}
where $\sigma$ is the noise level. The probability density function has units of inverse amplitude times inverse radians and is written as $p(a,\phi)dad\phi$ to make it explicit that it is a differential probability density. Expressed in terms of the real and imaginary parts of $\tilde{a}=x+iy$, the probability density function is a bivariate Gaussian distribution
\begin{equation}
p(x,y)dxdy = \frac{1}{2\pi\sigma^2}\exp\left[-\frac{x^2+y^2}{2\sigma^2}\right] dxdy.
\end{equation}
In the presence of a signal phasor with $\tilde{s}=s\cos{\delta}+is\sin{\delta}$ with amplitude s and phase $\delta$, in a noise background $\sigma$ the probability density function becomes 
\begin{equation}
p(x,y)dxdy = \frac{1}{2\pi\sigma^2}\exp\left[-\frac{\left(x-s\cos\delta\right)^2+\left(y-s\sin\delta\right)^2}{2\sigma^2}\right] dxdy.
\end{equation}

The likelihood function associated with the probability density function described is
\begin{equation}
-2\log L(x,y; s, \delta) = \frac{\left(x-s\cos\delta\right)^2+\left(y-s\sin\delta\right)^2}{\sigma^2}
\label{eq:ll1}
\end{equation}

\subsection{Likelihood Estimator of a Digitized Signal Observed by an Array of Antennas}
Let's assume we have an array of $N_A$ antennas indexed by $i$. If the signal at each antenna is digitized with sampling frequency $f_s$ with $N_s$ samples, then the frequency resolution is $\Delta f = f_s/N_s$. For uncorrelated noise, each frequency bin of width $\Delta f$ is an independent measurement indexed by $k$. For an array of antennas with effective height $h_{i,k}(\theta,\phi)$ observing an electric field $E_k$ (in one polarization), the antenna signal can be represented as $s_{i,k}=E_kh_{i,k}(\theta,\phi)\left(\cos\omega_k\tau(\theta,\phi)+i\sin\omega_k\tau(\theta,\phi)\right)$ where $\tau_{i}(\theta,\phi)$ is the geometrical delay of the signal $c\tau_{i}=\mathbf{R\cdot\hat{r}}$. For simplicity we have assumed that the overall electric field phase is at zero degrees and that the antenna effective height does not affect the phase other than by the geometric delay. Adding these dependencies is straightforward but we want to keep the following derivation as simple as possible.

Using Equation~\ref{eq:ll1}, the likelihood function for the array is given by
\begin{equation}
-2\log L(\{x_{i,k}\}, \{y_{i,k}\}; E,\omega,\theta,\phi) = \sum_{k=0}^{N_s/2}\sum_{i=1}^{N_A}
\frac{\left(x_{i,k}-E_{k}h_{i,k}\cos\omega_k\tau_{i}(\theta,\phi)\right)^2
+
\left(y_{i,k}-E_{k}h_{i,k}\sin\omega_k\tau_{i}(\theta,\phi)\right)^2}
{\sigma_{i,k}^2}
\end{equation}
where $\{x_{i,k}\}$ and $\{y_{i,k}\}$ are the measured real and imaginary parts for the phasors of each antenna indexed by $i$ at each frequency indexed by $k$. The sum over frequencies has $N_s/2+1$ independent contributions for a real digitized signal. 

The error for $\theta$ is estimated from the second derivative of the maximum likelihood estimator. The first derivative with respect to $\theta$ gives
\begin{equation}
\partial(-2\log L)/\partial\theta 
= 
\sum_{k=0}^{N_s/2}\sum_{i=1}^{N_A}
\frac{2E_{k}h_{i,k}\omega_k}{\sigma_{i,k}^2}\frac{\partial\tau_{i}}{\partial\theta}
\left[
x_{i,k}\sin\omega_k\tau_{i}-y_{i,k}\cos\omega_k\tau_{i}
\right]
\end{equation}

The second derivative with respect to $\theta$ gives
\begin{equation}
\partial^2(-2\log L)/\partial\theta^2
= 
\sum_{k=0}^{N_s/2}\sum_{i=1}^{N_A}
\frac{2E_{k}h_{i,k}}{\sigma_{i,k}^2}\omega_k
\left[
\omega_k\left(\frac{\partial\tau_i}{\partial\theta}\right)^2
\left(x_{i,k}\cos\omega_k\tau_{i}
+y_{i,k}\sin\omega_k\tau_{i}\right)
+
\left(\frac{\partial^2\tau_i}{\partial\theta^2}\right)
\left(x_{i,k}\sin\omega_k\tau_{i}-y_{i,k}\cos\omega_k\tau_{i}
\right)
\right]
\end{equation}
In the limit where the data approaches the modeled values $x_{i,k}\to E_{k}h_{i,k}\cos\omega_k\tau_i$ and $y_{i,k}\to E_{k}h_{i,k}\sin\omega_k\tau_i$
\begin{equation}
\partial^2(-2\log L)/\partial\theta^2
= 
2\sum_{k=0}^{N_s/2}\sum_{i=1}^{N_A}
\left(\frac{E_{k}h_{i,k}\omega_k}{\sigma_{i,k}}\frac{\partial\tau_{i}}{\partial\theta}\right)^2
\end{equation}
Note that $E_{k}h_{i,k}/\sigma_{i,k}$ is the signal to noise ratio ($snr_{i,k}$) at the antenna $i$ at frequency bin corresponding to $\omega_k$. 

For the purpose of illustration, let us assume we have a collinear array so that $\tau_{i}=(R_i/c) \cos\theta_{i}$, where $R_i$ is the distance to the origin and $\theta_i$ is the angle between $\mathbf{R}$ and $\mathbf{\hat{r}}$. Let us also assume that only one frequency bin has $snr_{i,k}\neq0$. The expression then becomes 
\begin{equation}
\partial^2(-2\log L)/\partial\theta^2
= 
2\sum_{i=1}^{N_A}
\left(snr_{i} \frac{2\pi}{\lambda}R\sin\theta_{i}\right)^2
\end{equation}
We estimate the angular error $\sigma_\theta$ using the relation $\sigma^{-2}_\theta=(1/2)\partial^2(-2\log L /\partial\theta^2)$. Note that the error is not arbitrarily small given the choice of a distant origin. If this is the case, all the angles $\theta_{i}$ will be small compensating for the choice of large $R_i$'s. Let us assume that a signal is incident orthogonal to the collinear array axis. If we set the origin at the location of one of the antennas, say the $N_A$th one, then its contribution drops out of the sum because $R_{N_A}=0$. Then we have
\begin{equation}
\sigma_{\theta}
= 
\frac{\lambda/2\pi}{\sqrt{\sum_{i=1}^{N_A-1}
snr_{i}^2 R_{i}^2}}
\end{equation}
Assuming that $snr_{i}=snr$ is the same for each antenna
\begin{equation}
\sigma_{\theta}
= 
\frac{\lambda/2\pi}{snr\sqrt{\sum_{i=1}^{N_A-1}R^2_{i}}}
\end{equation}
For the ANITA geometry, the antenna beam pattern is wide enough so that there are several vertical baseline pairs contributing to the elevation error. In this case there is a fixed vertical distance giving an improvement of roughly $\sqrt{N_A-1}$. The error can be calculated more rigorously using the effective heights of the antenna which make the channel to channel $snr$ vary.
% It is interesting to note that if the array consists of a large number of $N_A$ collinear antennas, placed in even steps of length $D$, then the improvement factor is $(N_{A}-1)^{3/2}$. 

In general, the digitization with $N_s$ records is sensitive to $N_s/2+1$ independent frequencies. Then we have
\begin{equation}
\partial^2(-2\log L)/\partial\theta^2
= 
\sum_{k=0}^{N_s/2}
\sum_{i=1}^{N_A}
\left(snr_{i,k} 2\pi\frac{R}{\lambda_k}\sin\theta_{i}\right)^2
\end{equation}

Let us assume that there is a relatively large number of frequency bins $N_f$, spanning indices $k_0$ to $k_1$, for which the signal to noise ratio $snr_{i,k}=snr$ is constant and non-zero. The error estimate is

\begin{equation}
\sigma_{\theta}
= \frac{c}{2\pi snr} 
\frac{1}{\sqrt{\sum_{k=k_0}^{k_1} f_{k}^2}}
\frac{1}{\sqrt{\sum_{i=1}^{N_A-1}R_{i}^2\sin^2\theta_i}}
\end{equation}

Let us evaluate $\sqrt{\sum_{k=k_0}^{k_1}f_{k}^2}$. For a digitized signal with $\Delta f$ we have $f_k =  k \Delta f$. For a large number of frequency bins the sum results in $\sum_{k=k_0}^{k_1} k^2 \approx (k_1^3-k^2_0)/3$. Factorization gives $k_1^3-k^2_0=(k_1-k_0)(k^2_1+k^2_0)$. The first term is the number of frequencies with non-zero SNR $(k_1-k_0)=N_f$. The second term gives $\Delta f^2 (k_1^2+k_0^2)/(3c^2)\approx1/\lambda^2$ for the central frequency $\lambda=2c/((k_1+k_0)\Delta f)$ Together these give
\begin{equation}
\sigma_{\theta}
\approx 
\frac{\lambda}{2\pi snr \sqrt{N_f}} 
\frac{1}{\sqrt{\sum_{i=1}^{N_A}R^2_{i}\sin^2\theta_i}}
\end{equation}
Thus, the direction error is reduced by a factor of $\sqrt{N_f}$. Intuitively, this result can be interpreted as each independent frequency providing an independent interferometric estimate of the incident angle $\theta$.

\subsection{Likelihood Estimator of an Interferometric Array}
A similar likelihood analysis can be performed on the cross-correlation phasor defined as the product $a_{i}a_{j}^{*}$ of two phasors. The cross-correlation phasor for a signal incident on the interferometric array is given by $E_k^2 h_{i,k}h_{j,k}\left[\cos\omega_k(\tau_i-\tau_j)+i\sin\omega_k(\tau_i-\tau_j)\right]$. The advantage of this approach is that any ambiguities due to choice of coordinate system vanish, since only delay differences between antenna pairs are counted. However, the estimation of the cross-correlation phasor noise is non-trivial and the accounting of independent contributions has additional complications. A maximum likelihood analysis technique along these lines has been developed for interferometric observations of the cosmic microwave background~\cite{Zhang_2012}. This approach is currently being developed as a potential improvement on the interferometric analysis presented on this paper and for a more rigorous accounting of errors. This will be the subject of a future publication. 


\bibliographystyle{elsarticle-num}
\bibliography{<your-bib-database>}

\begin{thebibliography}{00}

% \bibitem{goodman_1985} J. W. Goodman, \textit{Statistical Optics}, (John Wiley and Sons, 1985).
\bibitem{askaryan_1962} G. A. Askaryan, JETP 14, 441 (1962); JETP 21, 658 (1965).
% 
\bibitem{saltzberg_2001} D. Saltzberg et al., Phys. Rev. Lett., 86, 2802, (2001).
% 
\bibitem{ZHS92} E. Zas, F. Halzen, and T. Stanev, Phys. Rev. D 45, 362-376 (1992).
% 
\bibitem{gorham_2005} P.W. Gorham et al., Phys.Rev. D, 72, 023002, (2005) 
% 
\bibitem{gorham_2007} ANITA Collaboration: P.W. Gorham, et al., Phys. Rev. Lett. 99, 171101, (2007).
% 
\bibitem{rice_2006} I. Kravchenko, et al., Phys. Rev. D 73, 082002, (2006).
%
\bibitem{ARA} P. Allison, et al., Astropart. Phys. 35,  457–477, (2012).
%
\bibitem{GLUE} P. W. Gorham, et al., Phys. Rev. Lett. 93, 041101 (2004) 
% 
\bibitem{Lunaska} C. W. James, et al., Phys. Rev. D, 81, 042003 (2010) 
%
\bibitem{gorham_2009a} ANITA Collaboration: P.W. Gorham et al., Astropart. Phys. 32, 10-41, (2009)
%
\bibitem{falcke_2003} Falcke, H. and  Gorham, P., Astropart. Phys. 19, 477-494 (2003).
% 
\bibitem{suprun_2003} Suprun, D. A., Gorham, P. W., and Rosner, J. L., Astropart. Phys. 20 157-168, (2003).

% 
\bibitem{jelley_1965} Jelley, J. V. et al., Nature 205, 327-328 (1965).
% 
\bibitem{cr_1} Porter, N. A., Long, C. D., McBreen, B., Murnaghan, D. J. B. and Weekes, T. C., Phys. Lett. 19, 415-417 (1965).
% 
\bibitem{cr_2} Vernov, S. N., Abrosimov, A. T., Volovik, V. D., Zalyubovskii, I. I. and Khristiansen, G. B., Pis’ma v ZhETF 5, 157-162 (1967). [Sov. Phys. JETP Letters 5, 126-130 (1967)]
% 
\bibitem{cr_3} Barker, P. R., Hazen, W. E., and Hendel, A. Z., Phys. Rev. Lett. 18, 51-54 (1967).
% 
\bibitem{cr_4} Fegan, D. J. and Slevin, P. J., Nature 217, 440-441 (1968).
% 
\bibitem{cr_5} Hazen, W. E., Hendel, A. Z., Smith, H., and Shah N. J., Phys. Rev. Lett. 22, 35-37 (1969).
% 
\bibitem{cr_6} Hazen, W. E., Hendel, A. Z., Smith, H., and Shah N. J., 24, 476-479 (1970).
% 
\bibitem{cr_7} Spencer, R. E., Nature 222, 460-461 (1969).
% 
\bibitem{cr_8} Fegan, D. J. and Jennings, D. M., Nature 223, 722-723 (1969).
% 
\bibitem{cr_9} Allan, H. R., Progress in Elementary Particles and Cosmic Ray Physics, 10, edited by Wilson, J. G. and Wouthuysen S. G. (North-Holland, Amsterdam, 1971), 171-304, and references therein.
% 
\bibitem{cr_10} Ardouin, D. et al., Astropart. Phys. 31, 192-200 (2009).
% 
\bibitem{cr_11} Nehls, S. et al., Nucl. Instrum. Meth. A589, 350-361 (2008).
% 
\bibitem{cr_12} LOPES Collaboration, W. D. Apel et al., Astropart. Phys. 32, 294-303 (2010).
% 
\bibitem{hoover_2010} ANITA Collaboration: S. Hoover et al., Phys. Rev. Lett. 105, 151101 (2010). 
% 
\bibitem{TMS.1986} A.R. Thompson, J.M. Moran, G.W. Swenson, Interferometry and Synthesis in Radio Astronomy, John Wiley \& Sons, 1986
% 
\bibitem{VLBI98} O.J. Sovers, J.L. Fanselow, C.S. Jacobs, Rev. Mod. Phys. 70, 1393-1454 (1998)
% 
\bibitem{gorham_2010} ANITA Collaboration: P.W. Gorham et al., Phys. Rev. D 82, 022004 (2010); Phys. Rev. D 85, 049901(E) (2012)
%
\bibitem{Falcke2005} Falcke H., et al., Nature 435, 313-316 (2005) 
%
\bibitem{Miocinovic_2006} Mio\v{c}inovi\'c, P., et al., Phys. Rev. D 74, 043002 (2006) 
% 
\bibitem{ARZ_2011} J. Alvarez-Mu\~niz, A. Romero-Wolf, and E. Zas, Phys. Rev. D 84, 103003 (2011) 
% 
\bibitem{goodman_1985} J. W. Goodman, \textit{Statistical Optics}, John Wiley and Sons, 1985.
% 
\bibitem{gorham_2009b} ANITA Collaboration: P.W. Gorham et al., Phys. Rev. Lett. 103 051103, (2009)
%
% \bibitem{mathews_1970} J. Mathews and R.L. Walker, \textit{Mathematical Methods of Physics}, W.A. Benjamin, 1970.
%
% \bibitem{varner_2007} G. Varner et al., Nucl. Instrum. Meth. A583, 447-460, (2007)
\bibitem{RadarSat} Liu, H., Jezek, K., Li, B., and Zhao, Z.. 2001. Boulder, CO:
National Snow and Ice Data Center. Digital media
%
\bibitem{Besson_2013} D. Besson et al.,``Antarctic Radio Frequency Albedo and Implications for Cosmic Ray Reconstruction", arXiv 1301.4423 (2013) 
\bibitem{Zhang_2012} L. Zhang, A. Karacki, P. Sutter et al., ``Maximum Likelihood Analysis of Systematic Errors in the Interferometric Observations of the Cosmic Microwave Background" , http://arxiv.org/abs/1209.2676 (2012)
\end{thebibliography}



\pagebreak

\end{document}